\newif\ifsubmode
\submodefalse

\newif\ifprintfig
\printfigtrue

\ifsubmode
  \documentstyle[12pt,aasms4,epsf]{article}
  \received{}
  \accepted{}
  \journalid{}{}
  \articleid{}{}
\else
  \documentstyle[11pt,aaspp4,epsf]{article}
  \slugcomment{{\it Astronomical Journal, submitted May 2001.}}
\fi

\lefthead{van der Marel \& Cioni}
\righthead{Magellanic Cloud Structure I}

\newcommand{\etal}{{et al.~}}

\newcommand{\lta}{\lesssim}
\newcommand{\gta}{\gtrsim}

\newcommand{\kpc}{\>{\rm kpc}}

\newcommand{\chour}{^{\rm h}\>}
\newcommand{\cmin}{^{\rm m}\>}

\begin{document}

\title{Magellanic Cloud Structure from Near-IR Surveys I:\\ 
The Viewing Angles of the LMC}

\author{Roeland P.~van der Marel}
\affil{Space Telescope Science Institute, 3700 San Martin Drive, 
       Baltimore, MD 21218}

\author{Maria-Rosa L.~Cioni}
\affil{Sterrewacht Leiden, Postbus 9513, 2300 RA Leiden, 
       The Netherlands}



\ifsubmode\else
\clearpage\fi


\ifsubmode\else
\baselineskip=14pt
\fi


\begin{abstract}
We present a detailed study of the viewing angles of the LMC disk
plane. We find that our viewing direction differs considerably from
the commonly accepted values, which has important implications for the
structure of the LMC.

The discussion is based on an analysis of spatial variations in the
apparent magnitude of features in the near-IR color-magnitude diagrams
extracted from the DENIS and 2MASS surveys. Sinusoidal brightness
variations with a peak-to-peak amplitude of $\sim 0.25$ mag are
detected as function of position angle. The same variations are
detected for AGB stars (using the mode of their luminosity function)
and for RGB stars (using the tip of their luminosity function), and
these variations are seen consistently in all of the near-IR
photometric bands in both DENIS and 2MASS data. The observed spatial
brightness variations are naturally interpreted as the result of
distance variations, due to one side of the LMC plane being closer to
us than the opposite side. There is no evidence that any complicating
effects, such as possible spatial variations in dust absorption or the
age/metallicity of the stellar population, cause large-scale
brightness variations in the near-IR at a level that exceeds the
formal errors ($\sim 0.03$ mag). The best fitting geometric model of
an inclined plane yields an inclination angle $i = 34.7^{\circ} \pm
6.2^{\circ}$ and line-of-nodes position angle $\Theta =
122.5^{\circ}\pm 8.3^{\circ}$. The quoted errors are conservative
estimates that take into account the possible influence of systematic
errors; the formal errors are much smaller, $0.7^{\circ}$ and
$1.6^{\circ}$, respectively. There is tentative evidence for
variations of $\sim 10^{\circ}$ in the viewing angles with distance
from the LMC center, suggesting that the LMC disk plane may be warped.

Traditional methods to estimate the position angle of the line of
nodes have used either the major axis position angle $\Theta_{\rm
maj}$ of the spatial distribution of tracers on the sky, or the
position angle $\Theta_{\rm max}$ of the line of maximum gradient in
the velocity field, given that for a circular disk $\Theta_{\rm maj} =
\Theta_{\rm max} = \Theta$. The present study does not rely on the assumption 
of circular symmetry, and is considerably more accurate than previous
studies of its kind. We find that the actual position angle of the
line of nodes differs considerably from both $\Theta_{\rm maj}$ and
$\Theta_{\rm max}$, for which measurements have fallen in the range
$140^{\circ}$--$190^{\circ}$. This indicates that the intrinsic shape
of the LMC disk is not circular, but elliptical. Paper~II of this
series explores the implications of this result through a detailed
study of the shape and structure of the LMC. The inclination angle
inferred here is consistent with previous estimates, but this is to
some extent a coincidence, given that also for the inclination angle
most previous estimates were based on the incorrect assumption of
circular symmetry.
\end{abstract}



\clearpage



\section{Introduction}
\label{s:intro}

The Large Magellanic Cloud (LMC) is our close neighbor. It plays a key
role in determinations of the cosmological distance scale (e.g., Mould
\etal 2000) and is used to study the presence of dark objects in the
Galactic Halo through microlensing (e.g., Alcock \etal 2000b; Lasserre
\etal 2000; Udalski \etal 1999). It is also of fundamental importance
for all studies of stellar populations and the interstellar medium
(see the book on the Magellanic Clouds by Westerlund 1997). For all
these applications it is essential to have a good understanding of the
overall structure of the LMC as a galaxy, which is the topic of the
present series of papers.

The generally accepted consensus is that the LMC is a disk galaxy with
an approximately planar geometry (the evidence for this is reviewed in
Section~\ref{ss:planar}). The most basic parameters in our
understanding of the LMC are therefore the angles that describe the
direction from which we are viewing the LMC plane: the inclination
angle $i$ and the position angle $\Theta$ of the line of nodes (the
intersection of the galaxy plane and the sky plane). It is remarkable
that our knowledge of these parameters is only quite rudimentary,
given that the LMC is easily visible with the naked eye and is many
times the size of the full moon. The previous work on the LMC viewing
geometry is summarized in Table~3.5 of Westerlund (1997). The quoted
results for $i$ and $\Theta$ easily span a range of $25^{\circ}$ each,
even if one restricts the discussion to the most reliable studies.

The large majority of all previous studies of the LMC viewing geometry
have estimated the viewing angles under the assumption that the LMC
disk is circular, using either the spatial distribution of tracers on
the sky (as reviewed in Section~\ref{ss:circphot}) or their kinematics
(as reviewed in Section~\ref{ss:circkin}). The obvious disadvantage of
this approach is that there is no evidence that the LMC is indeed
circular. To the contrary, all available evidence seems to indicate
that the LMC is a highly complicated system with none of the
characteristic symmetries that are often seen in spiral galaxies. Its
interaction with the Small Magellanic Cloud (SMC) and the Galactic
tidal field (e.g., Putman \etal 1998) may be partly responsible for
this. The LMC hosts a central bar that is offset from the center of
the outer isophotes; the HI rotation center coincides with neither
(Westerlund 1997). The principal axes of the HI velocity field are not
perpendicular, and the zero-velocity curve twists by $\sim 20^{\circ}$
from small to large radii (Kim \etal 1998). So it seems reasonable to
worry about the accuracy of viewing angles inferred under the
assumption of circular symmetry.

A more robust way to determine the LMC viewing angles is to use
geometrical considerations without making assumptions about the
distribution or kinematics of tracers in the LMC plane. This is
possible, because the inclination of the LMC causes one side of it to
be closer to us than the other. As a result, tracers on one side of
the LMC should appear brighter than those on the other side. This is
not a subtle effect, but should amount to $\sim 0.2$--$0.4$ mag for
the viewing angles that are commonly quoted in the literature (as
discussed in Section~\ref{s:theory}). This method does not rely on
absolute distances or magnitudes, which are notoriously difficult to
estimate, but only on relative distances or magnitudes. What has been
lacking most is a large enough sample of stars to apply the method
to. Cepheids have yielded results, but so far only with large error
bars (as reviewed in Section~\ref{ss:cepheids}).

Large-area digital surveys such as the Magellanic Cloud Photometric
Survey (Zaritsky, Harris \& Thompson 1997), DENIS (Epchtein \etal
1997) and the Two Micron All Sky Survey (2MASS; e.g., Skrutskie 1998)
are now providing important new tools to analyze the LMC structure. In
particular, the LMC viewing geometry can be constrained by studying
how the apparent magnitude of well-defined features in the
color-magnitude diagram (CMD) varies as a function of position in the
LMC. A similar technique has been used to study the inclination of the
bar in the Milky Way (e.g., Stanek \etal 1994). Weinberg \& Nikolaev
(2000) applied this technique to the LMC, using Asymptotic Giant
Branch (AGB) stars selected from the 2MASS survey, but they did not
pursue this issue in as much detail as we do here (as discussed in
Section~\ref{ss:AGB2MASS}).

In this first paper of a series on LMC structure we analyze the LMC
viewing angles by studying how the characteristic apparent magnitudes
of AGB and Red Giant Branch (RGB) stars vary with position in the
LMC. The analysis is restricted to stars in the outer parts of the
LMC, $\geq 2.5^{\circ}$ from the LMC center. The analysis is based
primarily on the $I$, $J$ and $K_s$ band data from the DENIS survey,
and in particular the data collected in the DENIS Catalog towards the
Magellanic Clouds (DCMC; Cioni \etal 2000a). However, 2MASS data are
used as well. Section~\ref{s:theory} presents the theoretical basis of
the analysis. Coordinate systems are introduced, and it is derived how
brightness variations on the sky are related to the LMC viewing
geometry. Section~\ref{s:method} describes the method used for the
analysis of near-infrared (near-IR) CMDs. Section~\ref{s:datamodel}
discusses the technique for data-model comparison to obtain estimates
for the viewing angles from the data. Section~\ref{s:results} presents
the results of an analysis of the $I$-band magnitudes of AGB stars
selected by $I-J$ color. Section~\ref{s:difbands} addresses the extent
to which the results of this analysis depend on the photometric
band(s) used for the analysis or the color
selection. Section~\ref{s:TRGB} compares the spatial variations in the
brightness of the tip of the RGB (TRGB) with the results obtained for
AGB stars. Section~\ref{s:varm} studies variations in the
characteristic magnitudes of AGB stars with distance from the LMC
center. Section~\ref{s:modassump} provides a detailed assessment of
possible sources of systematic error, including spatial variations in
dust absorption or the age/metallicity of the stellar
population. Also, results from DENIS and 2MASS data are compared to
assess the influence of possible systematic errors in the data.
Section~\ref{s:warptwist} studies the dependence of the viewing angles
on the distance from the LMC center, which constrains warps and twists
of the LMC disk plane. Section~\ref{s:comparison} discusses how the
inferred viewing angles compare to the results of previous authors.
Section~\ref{s:indpos} addresses the implications for the relative
positions and distances of some well-studied objects, supernova
SN1987A and two eclipsing binaries, with respect to the LMC
center. Section~\ref{s:conc} presents concluding remarks.
Appendix~\ref{s:AppA} discusses the photometric accuracy of the DCMC,
and discusses improvements made to the photometric zeropoint
calibrations of Cioni \etal (2000a).

Paper~II of this series (van der Marel 2001a) studies the intrinsic
shape and structure of the LMC, by combining the observed number
density distribution of stars inferred from the 2MASS and DENIS
surveys with the viewing angles inferred here. Paper~III (van der
Marel 2001b) addresses the question whether the structures in the
central $2.5^{\circ}$, including the LMC bar, lie in the same plane as
that defined by the outer disk.

\section{Theoretical Framework}
\label{s:theory}

\subsection{Angular Coordinates}
\label{ss:angdist}

The position of any point in space is uniquely determined by its right
ascension (RA) and declination (DEC) on the sky, $(\alpha,\delta)$,
and its distance $D$. A particular point ${\cal O}$ with coordinates
$(\alpha_0,\delta_0,D_0)$ is taken to be the origin of the
analysis. This point will typically be chosen to be the center of the
LMC, but the formulae that follow are valid more generally.

Angular coordinates $(\rho,\phi)$ are defined on the celestial sphere,
where $\rho$ is the angular distance between the points
$(\alpha,\delta)$ and $(\alpha_0,\delta_0)$, and $\phi$ is the
position angle of the point $(\alpha,\delta)$ with respect to
$(\alpha_0,\delta_0)$. In particular, $\phi$ is the angle at
$(\alpha_0,\delta_0)$ between the tangent to the great circle on the
celestial sphere through $(\alpha,\delta)$ and $(\alpha_0,\delta_0)$,
and the circle of constant declination $\delta_0$. By convention,
$\phi$ is measured counterclockwise starting from the axis that runs
in the direction of decreasing RA at constant declination $\delta_0$.

The cosine rule of spherical trigonometry (e.g., Smart 1977) can be
used to derive that
\begin{equation}
\cos \rho = \cos \delta \cos \delta_0 \cos (\alpha - \alpha_0) +
            \sin \delta \sin \delta_0 .
\label{rhodef}
\end{equation}
Also, the sine rule of spherical trigonometry can be used to
show that
\begin{equation}
\sin \rho \cos \phi = - \cos \delta \sin (\alpha - \alpha_0) ,
\label{phidef1}
\end{equation}
and the so-called analogue formula implies that
\begin{equation}
\sin \rho \sin \phi =  \sin \delta \cos \delta_0 -
        \cos \delta \sin \delta_0 \cos (\alpha - \alpha_0) .
\label{phidef2}
\end{equation}
These formulae uniquely define $(\rho,\phi)$ as function of
$(\alpha,\delta)$, for a fixed choice of the origin ${\cal O}$.

It is often useful to plot observations on the celestial sphere
on a flat piece of paper. This requires transformation equations from
$(\alpha,\delta)$ to a cartesian coordinate system $(X,Y)$. We will
use
\begin{equation}
  X(\alpha,\delta) \equiv \rho \cos \phi , \qquad
  Y(\alpha,\delta) \equiv \rho \sin \phi .
\label{projdef}
\end{equation}
This so-called `zenithal equidistant projection' provides just one of
the many possible ways of projecting a sphere onto a plane (see, e.g.,
Calabretta 1992).

\subsection{Relative Distances and Magnitudes}
\label{ss:reldistmag}

A cartesian coordinate system $(x,y,z)$ is introduced that has its
origin at ${\cal O}$, with the $x$-axis anti-parallel to the RA axis,
the $y$-axis parallel to the declination axis, and the $z$-axis
towards the observer. This yields the following transformations:
\begin{eqnarray}
\label{xyztransf}
  x & = & D \sin \rho \cos \phi , \nonumber\\
  y & = & D \sin \rho \sin \phi , \\
  z & = & D_0 - D \cos \rho . \nonumber
\end{eqnarray}
If necessary, these equations can be expressed in terms of
$(\alpha,\delta)$ using equations~(\ref{rhodef})--(\ref{phidef2}) (see
also Appendix A of Weinberg \& Nikolaev 2000).

A second cartesian coordinate system $(x',y',z')$ is introduced that
is obtained from the system $(x,y,z)$ by counterclockwise rotation
around the $z$-axis by an angle $\theta$, followed by a clockwise
rotation around the new $x'$-axis by an angle $i$. With this
definition, the $(x',y')$ plane is inclined with respect to the sky by
the angle $i$ (with face-on viewing corresponding to $i=0$). The angle
$\theta$ is the position angle of the line of nodes (the intersection
of the $(x',y')$-plane and the $(x,y)$-plane of the sky), measured
counterclockwise from the $x$-axis. In practice, $i$ and $\theta$ will
be chosen such that the $(x',y')$ plane coincides with the plane of
the LMC disk, but the formulae that follow are valid more generally.
The transformations between the $(x',y',z')$ and $(x,y,z)$ coordinates
are:
\begin{eqnarray}
\label{xyzprimetransf}
x' & = & x \cos \theta + 
          y \sin \theta , \nonumber\\
y' & = & - x \sin \theta \cos i +
         y \cos \theta \cos i -
         z \sin i , \\
z' & = & - x \sin \theta \sin i +
         y \cos \theta \sin i +
         z \cos i . \nonumber
\end{eqnarray}
Substitution of equation~(\ref{xyztransf}) yields 
\begin{eqnarray}
\label{xyzprimetransfpolar}
x' & = & D \sin \rho \cos (\phi - \theta) , \nonumber\\
y' & = & D [\sin \rho \cos i \sin (\phi - \theta) + \cos \rho \sin i] -
         D_0 \sin i , \\
z' & = & D [\sin \rho \sin i \sin (\phi - \theta) - \cos \rho \cos i] +
         D_0 \cos i . \nonumber
\end{eqnarray}
Here the elementary rules $\cos(\phi-\theta) =
\cos \phi \cos \theta + \sin \phi \sin \theta$ and 
$\sin(\phi-\theta) = \sin \phi \cos \theta - \cos \phi \sin \theta$
were used, which follow from the complex identity $e^{i(\phi-\theta)}
= e^{i\phi} e^{-i\theta}$.
 
One is interested in the distance $D$ of points in the $(x',y')$
plane, as function of the position $(\rho,\phi)$ on the sky. The
points in this plane have $z' = 0$, so that
equation~(\ref{xyzprimetransfpolar}) yields:
\begin{equation}
D / D_0 = \cos i \> / \> 
             [\cos i \cos \rho -
              \sin i \sin \rho \sin(\phi-\theta) ]  .
\label{Dsimpler}
\end{equation}
This general equation simplifies in a number of special cases. If one
considers points along the line of nodes, or if the system is viewed
face on, then
\begin{equation}
D / D_0 = 1 / \cos \rho \qquad 
  ({\rm for}\>\> \phi = \theta ,\>\>
   \phi = \theta + 180^{\circ} ,\>\>
   {\rm or}\>\> i=0) . 
\label{Dnodesorfaceon}
\end{equation}
On the other hand, if one considers points along a line that is
perpendicular to the line of nodes then
\begin{equation}
D / D_0 = \cos i / \cos (i \pm \rho) \qquad
  ({\rm for}\>\> \phi = \theta \pm 90^{\circ}) ,
\label{Dmaxandmin}
\end{equation}
which uses the elementary rule that $\cos(i \pm \rho) = \cos i \cos
\rho \mp \sin i \sin \rho$. For a fixed angular distance $\rho$ from the 
origin ${\cal O}$, the points with $\phi = \theta \pm 90^{\circ}$ are
the ones for which $D / D_0$ reaches its maximum and minimum values,
respectively. For small angular distances $\rho$ one can expand
equation~(\ref{Dsimpler}) using a Taylor expansion to obtain that
\begin{equation}
D / D_0 = 1 + \rho \tan i \sin(\phi-\theta) + {\cal O}(\rho^2) , \qquad
  (\rho \>\> {\rm in}\>\> {\rm radians,}\>\> \rho \ll 1) .
\label{Dtaylor}
\end{equation}

In practice one does not directly have access to distances, but one
does have access to stellar magnitudes. Consider identical objects at
positions $(\alpha,\delta)$ and $(\alpha_0,\delta_0)$. The apparent
magnitudes of these objects will differ by
\begin{equation}
  \mu \equiv m - m_0 = 5 \log (D/D_0) .
\label{magdif}
\end{equation}
The same will hold for the average magnitudes of groups of objects
with identical properties. In general, $\mu$ can be evaluated as
function of $(\rho,\phi)$ (i.e., position on the sky) using
equation~(\ref{Dsimpler}), for any given viewing angles
$(i,\theta)$. For small angular distances $\rho$, one can use
equation~(\ref{Dtaylor}) and make a further Taylor expansion of the
logarithm to obtain
\begin{equation}
\mu = 
   \Big ( { {5\,\pi} \over {180\ln 10} } \Big ) \> 
     \rho \tan i \sin(\phi-\theta) 
     + {\cal O}(\rho^2) , \qquad
     (\rho \>\> {\rm in}\>\> {\rm degrees,}\>\> \rho \ll 180/\pi) ,
\label{dmagtaylor}
\end{equation}
where the angular distance $\rho$ is now expressed in degrees, to make
it easier to assess the size of the magnitude difference for realistic
situations. The constant in the equation is $(5\pi) / (180\ln 10) =
0.038$ magnitudes. Hence, the magnitude at fixed $\rho$ has an
approximately sinusoidal variation with position angle $\phi$, with
amplitude $0.038 \> \rho \tan i$. Stars in the LMC can be traced to
radii $\rho$ of 5--10 degrees from the center, and previous analyses
have suggested that $i \approx 45^{\circ}$ (so that $\tan i \approx
1$; e.g., Westerlund 1997). Hence, in the LMC one expects
distance-related magnitude variations up to several tenths of a
magnitude (much larger than typical observational errors). At fixed
$\rho$, the magnitude variation $\mu$ always reaches its extrema at
angles perpendicular to the line of nodes, $\phi = \theta
\pm 90^{\circ}$. Figure~\ref{f:maxminvar} shows $\mu$
as function of $\rho$ along this line, for different values of the
inclination $i$, as calculated from equations~(\ref{Dmaxandmin})
and~(\ref{magdif}). For comparison, the heavy long-dashed line shows
the linear approximation given by equation~(\ref{dmagtaylor}), for
$i=40^{\circ}$.

\subsection{Position angles}
\label{ss:posang}

The usual method of measuring position angles in astronomy is to
measure counterclockwise, starting from the North. By contrast, the
angles $\phi$ and $\theta$ defined above are measured counterclockwise
starting from the West. It is therefore useful to define
\begin{equation}
  \Phi = \phi - 90^{\circ} , \qquad 
  \Theta = \theta - 90^{\circ} ,
\label{posangfromnorth}
\end{equation}
which are the position angle of a point in the LMC and the position
angle of the line-of-nodes, respectively, now measured with the usual
astronomical convention. 

The description of the LMC viewing geometry using the angle $\Theta$
can lead to some confusion, since the angle $\Theta$ is defined modulo
$2 \pi$, whereas the line-of-nodes is a line, and can therefore be
described by two different position angles that differ by $\pi$. The
definition used here corresponds to the usual convention (e.g.,
Westerlund 1997), by which the quantity referred to as the `position
angle of the line-of-nodes' (i.e., $\Theta$) is defined such that
points in the direction of position angle $\Theta_{\rm near} \equiv
\Theta - 90^{\circ}$ are closer to the observer than those in the
direction of position angle $\Theta_{\rm far} \equiv \Theta +
90^{\circ}$.

\section{Methodology to Constrain the LMC Structure from Near-IR Photometry}
\label{s:method}

\subsection{The DENIS Data}
\label{ss:DCMC}

We study the apparent magnitude of well-defined features in the CMD as
function of position in the LMC to determine the viewing angles using
the formulae of Section~\ref{s:theory}. Large-scale digital surveys
such as the Magellanic Cloud Photometric Survey (Zaritsky
\etal 1997), DENIS (Epchtein \etal 1997) and 2MASS (e.g., Skrutskie 1998)
are now yielding catalogs with up to millions of stars, which are
ideally suited for a study of this nature.

In the present paper the discussion is restricted mostly to the
near-IR data from the DENIS survey, and in particular to the data
collected in the DENIS Catalog towards the Magellanic Clouds (DCMC;
Cioni \etal 2000a). We used the latest version, which includes the
data for the few scan strips missing from the first release. The
resulting catalog has complete coverage over the LMC area with
$4\chour 06\cmin \leq$ RA $\leq 6\chour 47\cmin$ and $-77^{\circ}
\leq$ DEC $\leq -61^{\circ}$ (see Figure~\ref{f:imagegrid} below). It
contains stellar magnitudes of sources detected in at least two of the
three DENIS bands ($I$, $J$ and $K_S$). For the present analysis all
sources in the catalog were used. Sources with non-optimal values of
any of the DCMC data-quality flags were not excluded to optimize the
statistics of the sample (several tests were performed to verify that
this does not degrade the accuracy of the final results). An improved
calibration of the photometric zeropoints for the individual DENIS
scan strips was performed. This significantly increased the accuracy
of the DCMC, as discussed in Appendix~\ref{s:AppA}. Systematic errors
in the resulting stellar magnitudes are believed to be no larger than
$\sim 0.02$ mag, i.e., much smaller than the expected distance-induced
magnitude variations (cf.~Figure~\ref{f:maxminvar}). The issue of
possible systematic errors is further addressed in
Section~\ref{ss:caterrors}, among other things by comparison to 2MASS
data.

\subsection{Near-IR Color Magnitude Diagrams}
\label{ss:cmd}
 
Figure~\ref{f:CMDs} shows the nine CMDs that can be generated from the
$I$, $J$ and $K_S$ data in the DCMC. The general features of these
near-IR CMDs of the LMC have been previously discussed by Cioni \etal
(2000a,b,c). Detailed discussions of the near-IR $(J-K_s,K_s)$ and
optical $(V-R,V)$ CMDs of the LMC have been presented by Nikolaev \&
Weinberg (2000) and Alcock \etal (2000a), respectively. We therefore
summarize here only the main features of the CMDs in
Figure~\ref{f:CMDs}, as relevant in the present context.

LMC stars in the RGB evolutionary phase are responsible for the
pronounced feature that extends downward, slanted somewhat to the
left, from the center of each CMD panel in Figure~\ref{f:CMDs}. The
horizontal bar at the right axis of each panel indicates the magnitude
of the Tip of the Red Giant Branch (TRGB), as determined by Cioni
\etal (2000c). LMC stars in the AGB evolutionary phase are responsible
for feature(s) at magnitudes that are brighter and redder than the
TRGB. There are two main types of AGB stars, namely the oxygen-rich
(O-rich) and the carbon-rich (C-rich) AGB stars. In the CMDs involving
the $J-K_s$ color (bottom panels in Figure~\ref{f:CMDs}) these
families separate into easily distinguishable features. The O-rich AGB
stars have an approximately constant $J-K_s$ color, and therefore
generate a feature that extends upwards almost vertically from the
TRGB. The C-rich AGB stars generally have a redder $J-K_s$ color than
the O-rich AGB stars, and generate the feature in the $J-K_s$ CMDs
extending to the right or top-right starting from the O-rich AGB
feature with minimum overlap. In some of the other CMDs in
Figure~\ref{f:CMDs} the two families of AGB stars occupy overlapping
regions of color-magnitude space, e.g., in the $(I-J,I)$ and $(I-J,J)$
diagrams. The relatively blue stars on the left-side of each CMD that
lie in vertical features extending upward to very bright magnitudes
are in large majority Galactic foreground stars.

The goal in the present context is to construct CMDs for different
spatial regions in the LMC, and to compare them. If the regions are at
different distances then {\it all} features in the CMD due to stars in
the LMC will shift up or down in magnitude by the same amount. So the
first order of business is to determine on a purely empirical basis
what vertical magnitude shifts there are between the CMDs at different
spatial positions. In principle one could use the full two-dimensional
data in each CMD to obtain these magnitude shifts, for example by
cross-correlation of the CMDs at different positions. This requires
only that one restricts the cross-correlation to a region of the CMD
that contains little if any Galactic foreground contamination and that
contains well-defined features due to LMC stars. Nonetheless, this may
not be entirely trivial to implement for a dataset of discrete
points. The more straightforward approach is therefore to analyze
one-dimensional luminosity functions (LFs) that are obtained upon
projection of a CMD along its color axis. In this projection it is
sometimes useful to constrain the LF to stars with a particular range
of colors, to restrict the analysis to a more homogeneous set of stars
(e.g., stars in a similar evolutionary phase) or to remove foreground
stars.

\subsection{Extraction of Luminosity Functions}
\label{ss:AGBLF}
 
Near-IR luminosity functions for the LMC contain several features that
can be used to determine a characteristic magnitude. Of particular use
are the features due to RGB and AGB stars. One prominent feature is
the TRGB, which is a discontinuity in the LF of RGB stars. The TRGB
magnitude can be determined with high accuracy for the LMC as a whole,
by searching for a peak in the first or the second derivative of the
LF (Cioni \etal 2000c). However, one does need a very large sample of
stars to obtain a high signal-to-noise ratio ($S/N$) even after
differentiation of the LF. By contrast to RGB stars, AGB stars do not
produce a well-defined discontinuity in the LF. However, they do
produce a well-defined peak (e.g., see Figure~\ref{f:histograms}, to
be discussed below). Hence, the most prominent LF feature due to AGB
stars is not their maximum brightness (as for the RGB), but the mode
of their magnitudes (the magnitude at which the LF has its
maximum). This AGB modal magnitude can be determined much more
accurately (for a fixed area of the sky) than the TRGB magnitude,
because no differentiation of the LF is required. Since the aim of the
present paper is to subdivide the LMC into different spatial regions
(possibly with limited numbers of stars), the TRGB method is not the
optimal choice in the present context. Nonetheless, it will be used in
Section~\ref{s:TRGB} as a consistency check on the results obtained
from AGB stars.

The accuracy with which the AGB modal magnitude can be determined
depends on the photometric band of the magnitudes under study, and the
color-selection applied to the AGB stars. The determining factors are
the number of stars in the LF peak (the more stars, the more accurate
the result) and the width of the LF peak (the narrower the peak, the
more accurate the result); see Section~\ref{ss:LFanalysis} below for
quantitative details. In the present context it was found that the
best results are obtained when the analysis is based on the $(I-J,I)$
CMD. In this diagram the entire AGB feature, including both O-rich and
C-rich AGB stars, is almost horizontal.  Hence, the $I$ magnitude LF
of stars selected by $I-J$ color makes a very useful `standard
candle'. For the primary analysis of the present paper the color
selection criterion $1.6 \leq I-J \leq 2.0$ was used (indicated by
vertical lines in the top left panel of Figure~\ref{f:CMDs}). The
lower-limit was chosen to avoid a significant contribution of RGB
stars to the LF. The upper limit was chosen based on tests that showed
that inclusion of the relatively small number of stars with $I-J >
2.0$ did not lead to a noticeable improvement in the accuracy of the
final results. Figure~\ref{f:histograms} shows the $I$-band LF of
stars thus selected from the DCMC. The peak due to AGB stars has a
Gaussian width of $\sigma \approx 0.3$ mag.

While it was found that particularly accurate results are obtained by
studying the $I$-band LF of stars selected by $I-J$ color, the
presence of a pronounced AGB peak in the LF is not uniquely obtained
only with this choice. The same color selection yields a peaked LF in
all three of the near-IR photometric bands, and the same is true for
selection based on $I-K$ or $J-K$ color. In Section~\ref{s:difbands}
it is shown that the main results of the analysis are independent of
which approach is adopted.

\subsection{Analysis of AGB Luminosity Functions}
\label{ss:LFanalysis}

For the analysis of the DCMC data the sky area of the LMC was
subdivided into disjunct sectors (as described in
Section~\ref{ss:spatial} below). For each sector the LF histogram of
those stars in a fixed range of colors was extracted (in analogy with
Figure~\ref{f:histograms}). To quantify the magnitude of the mode of
the distribution a Gaussian is fit to the peak of the LF, and the mean
of the best-fitting Gaussian is adopted as an estimate of the mode.
Experiments with various algorithms show that neither the method of
binning the data to obtain the LF nor the method of fitting the
Gaussian make any significant difference on the final results of the
analysis. The results presented here were obtained by binning the
stellar magnitudes in bins of $0.15$ mag. The Gaussian fit was
subsequently performed by minimizing a $\chi^2$ quantity that measures
the difference between the Gaussian model and the LF histogram over an
interval of size 1 mag around the peak.

The LFs extracted from the data are not entirely symmetric, and a
Gaussian fit may therefore not yield a completely unbiased estimate of
the true mode. However, this should not be very important in the
present context. If the shape of the LF is not strongly dependent on
position in the LMC, then the bias in the estimate of the mode should
be similar for different spatial positions. Such a spatially constant
bias will not affect the analysis, since one is only interested in the
{\it relative} distances and magnitude differences for different parts
of the LMC.

To estimate the formal measurement errors on the results of the
Gaussian fits we have performed Monte-Carlo simulations. In these
simulations stellar magnitudes are drawn from a Gaussian distribution,
for a given assumed number of stars $N$ and dispersion $\sigma$ of the
Gaussian magnitude distribution. These magnitudes are binned and
analyzed similarly to the real data. This procedure is repeated in
Monte-Carlo fashion, and the distribution of the inferred peak
magnitudes is then studied. The dispersion of this distribution
corresponds to the formal measurement error $\Delta m$ that one should
expect for the assumed $N$ and $\sigma$. From simulations with
different values of $N$ and $\sigma$ it was found that for the range
of values relevant to the analysis (namely $N \gta 100$ and $\sigma
\lta 0.5$) the formal error is well-described (to within $\sim 10$\%)
by the formula\footnote{Note that the order of magnitude of this
result makes immediate intuitive sense, since the formal error in the
average of $N$ measurements with RMS $\sigma$ equals $\sigma /
\sqrt{N}$. Note also that $\Delta m$ is much smaller than the adopted
binsize, which therefore does not limit the accuracy of the results.}
\begin{equation}
  \Delta m \approx 1.4 \> \sigma / \sqrt{N} =
  0.013 \> (\sigma / 0.3) (N/1000)^{-1/2}  .
\label{formalerror}
\end{equation}
To use this formula in practice one needs an estimate of the number of
stars $N$. For this we use the area under the best-fitting Gaussian.
This is better than to use the actual number of stars in the LF
histogram, since some of these stars have no influence on the fitting
of the LF peak. For the quantity $\sigma$ in
equation~(\ref{formalerror}) we use the dispersion of the best-fitting
Gaussian. The accuracy of the resulting formal measurement errors will
be discussed in Section~\ref{s:difbands}.

The random (noise) errors in the individual stellar magnitudes in the
catalog are quite negligible for the relatively bright AGB stars of
interest here: $\Delta I = 0.01$, $\Delta J = 0.01$ and $\Delta K_s =
0.03$ (see Appendix~\ref{s:AppA}). This is much smaller than the
intrinsic width of the peak of the LF, and these random errors
therefore do not affect the accuracy of the determination of the modal
magnitude.

To get a feeling for the magnitude scales involved, note that for the
histogram in Figure~\ref{f:histograms} one has $N \approx 13,400$
(i.e., approximately 4.5\% of all the stars in the DCMC that were
detected in all three of the DENIS bands). So if one divides the LMC
in 10 sectors with equal numbers of stars, then the modal AGB
magnitude of stars selected to have $1.6 \leq I-J \leq 2.0$ can be
determined with an error of $\Delta m \approx 0.011$ mag for each
sector. If one chooses a finer subdivision in 100 sectors then the
error goes up to $\Delta m \approx 0.036$ mag. The expected
peak-to-peak distance-induced magnitude variations are $\sim 10$ times
larger than this (cf.~Figure~\ref{f:maxminvar}). Hence, the available
statistics are more than adequate for a detailed study of the LMC
viewing angles.

\section{Formalism for Data-Model Comparison}
\label{s:datamodel}

\subsection{The Sky Grid}
\label{ss:spatial}

The analysis in the present paper is restricted to the outer parts of
the LMC, $\rho \geq 2.5^{\circ}$. The motivation for this is that some
previous work has suggested that structures in the inner parts of the
LMC ($\rho < 2.5^{\circ}$) may be decoupled from the outer LMC disk. A
clear hint in this direction is that the most prominent feature in the
central few degrees of the LMC, the `bar', is offset from the center
of the outer disk by $\sim 0.5^{\circ}$ (Westerlund 1997). It has
recently been suggested that the bar may also not be in the same plane
as the outer disk (Zhao \& Evans 2000). The second most prominent
feature in the central few degrees of the LMC is the 30 Doradus
complex. This region is a very strong source of UV radiation, yet the
HI gas disk of the LMC does not show a void in this part of the
sky. This indicates that the 30 Doradus complex cannot be in the plane
of the LMC disk (Luks \& Rohlfs 1992). The 30 Doradus region is also
the center of a separate velocity component seen in radio data (the
L-component; Luks \& Rohlfs 1992) for which absorption studies
indicate that it is behind the LMC disk (Dickey \etal 1994). Based on
these considerations the discussion here is restricted to the region
$\rho > 2.5^{\circ}$. Paper~III addresses the important question
whether the structures at $\rho < 2.5^{\circ}$ lie in the same plane
as determined here for the outer LMC disk.

The angular coordinates $(\rho,\Phi)$ defined in
Sections~\ref{ss:angdist} and~\ref{ss:posang} are used to divide the
region of the sky occupied by the LMC into disjunct sectors $A_{ln}$,
with
\begin{equation}
A_{ln} = \lbrace (\rho,\Phi) : 
            \rho_{l} \leq \rho < \rho_{l+1} \wedge
            \Phi_{n} \leq \Phi < \Phi_{n+1} \rbrace , \qquad
  l=1,\ldots,L, \qquad n=1,\ldots,N .
\label{Alndef}
\end{equation}
Here $L$ and $N$ are the number of radial and azimuthal bins. The
arrays $\lbrace \rho_{l} \rbrace$ and $\lbrace \Phi_{n} \rbrace$ mark
the radial and azimuthal grid boundaries, respectively. The spatial
grid adopted for the analysis is shown in
Figure~\ref{f:imagegrid}. The azimuthal grid is linearly spaced with
step size $\Delta \Phi = 360^{\circ} / N$. We chose to use $N=8$,
yielding wedges of opening angle $\Delta \Phi = 45^{\circ}$. The
radial grid was chosen to yield 4 rings, containing the radii $\rho$
in the range $(2.5^{\circ},3.4^{\circ})$, $(3.4^{\circ},4.4^{\circ})$,
$(4.4^{\circ},5.5^{\circ})$ and $(5.5^{\circ},6.7^{\circ})$,
respectively. The outer radius of the grid, $\rho = 6.7^{\circ}$ is
imposed by the spatial coverage of the DCMC. The origin ${\cal O}$ of
the grid was chosen at the position with RA = $5\chour 29\cmin$ and
DEC = $-69.5^{\circ}$, which corresponds roughly to the center of the
outer DCMC isoplets (cf.~Paper~II).

\subsection{Finding the Best-Fit Model}
\label{ss:viewdeterm}

To interpret the data it is assumed that the LMC resides in a thin
plane. In this case, equations~(\ref{Dsimpler}) and~(\ref{magdif}) can
be combined to provide model predictions $\mu(\rho,\Phi;i,\Theta)$ for
the magnitude variation as a function of position $(\rho,\Phi)$, given
viewing angles $(i,\Theta)$. Let $\mu_{ln} (i,\Theta)$ be the
corresponding model prediction integrated over the sector $A_{ln}$,
given by
\begin{equation}
  \mu_{ln} (i,\Theta) = 
      \int\!\!\int_{A_{ln}} \Sigma(\rho,\Phi) \mu(\rho,\Phi;i,\Theta) 
                \> {\rm d}A   \> \Big / \> 
      \int\!\!\int_{A_{ln}} \Sigma(\rho,\Phi) 
                \> {\rm d}A   ,
\label{areaintegral}
\end{equation}
where $\Sigma(\rho,\Phi)$ is the number density distribution on the
sky, and ${\rm d}A = \sin \rho \> {\rm d}\rho \> {\rm d}\Phi$ is an
infinitesimal surface area element.

The data-model comparison is characterized by the equations
\begin{equation}
  \mu_{ln} (i,\Theta) + m_0 = m_{ln} \pm \Delta m_{ln} , \qquad
  l = 1,\ldots,L ,\quad  n = 1,\ldots,N .
\label{fitequations}
\end{equation}     
Here $m_{ln}$ and $\Delta m_{ln}$ are an observed apparent magnitude
and its formal error, respectively, for sector $A_{ln}$ (the main
analysis uses AGB modal magnitudes as determined in
Section~\ref{ss:LFanalysis}, but equation~(\ref{fitequations}) is
equally valid for any other characteristic magnitude of the stellar
population, such the TRGB magnitude studied in
Section~\ref{s:TRGB}). The quantity $m_0$ is the apparent magnitude if
the stars of interest had been observed at the origin of the
coordinate system (the LMC center). This quantity is not known {\it a
priori}, and must therefore be obtained from a fit to the data. In
practice the best-fitting $(i,\Theta,m_0)$ are obtained by minimizing
the $\chi^2$ quantity
\begin{equation}
  \chi^2 \equiv 
     \sum_{l=1}^{L} \sum_{n=1}^{N} 
     \Bigl ( { {m_{ln} - \mu_{ln} (i,\Theta) - m_0} \over
               {\Delta m_{ln}} }
     \Bigr )^2 .
\label{chisqdef}
\end{equation}     

One is not forced to assume in the data-model comparison that
$(i,\Theta,m_0)$ are constant throughout the galaxy. Instead, if
$(i,\Theta,m_0)$ are allowed to be different for each radial ring on
the sky (fixed $l$), then one can search for variations in these
quantities as function of distance $\rho$ from the LMC center. We
consider both models in which $(i,\Theta)$ are constant, as well as
models in which $(i,\Theta)$ are allowed to vary as function of
$\rho$. The latter models provide constraints on possible warps and
twists in the LMC disk plane. The quantity $m_0$ was allowed to vary
as function of $\rho$ in all of the fits, to allow for possible radial
gradients in dust absorption or the stellar population mix.

\section{$I$-band Results for AGB Stars Selected by $I-J$ Color}
\label{s:results}

The data points in Figure~\ref{f:fits} show the results obtained from
the $I$-band LF of stars selected to have $1.6 \leq I-J \leq 2.0$.
Each panel corresponds to a different radial ring, with the innermost
ring in the top panel and the outermost ring in the bottom panel. The
position angle $\Phi$ is plotted along the abscissa, with each
datapoint corresponding to a different azimuthal sector. The quantity
$\mu$ along the ordinate (defined in eq.~[\ref{magdif}]) is the
difference between the AGB modal magnitude (inferred from the LF as
described in Section~\ref{ss:LFanalysis}) and the quantity $m_0$
(obtained from the model fit as described in
Section~\ref{ss:viewdeterm}). The curves in the figure show the best
fits to the data. Dashed curves show the fits when only a single
combination of the viewing angles $(i,\Theta)$ is allowed for all
radial rings, while solid curves show the fits when $(i,\Theta)$ are
allowed to be different for each radial ring. The corresponding models
will be loosely referred to as the best-fit `radially-constant' and
`radially-varying' models, respectively.

For the best-fit radially-varying model the RMS residual of the fit is
$0.027$ mag. This is $\sim 10$ times smaller than the peak-to-peak
amplitude of the azimuthal variations in $\mu$.  The RMS residual is
similar to the average of the formal errors in the $\mu$ measurements,
which is $0.029$ mag. The overall $\chi^2$ of the fit for the best-fit
radially-varying model is $30.1$ (for 32 datapoints), which indicates
that the fit to the data is good. If Gaussian random errors are
assumed to be the only relevant source of error (a considerable
oversimplification) then the $\chi^2$ is acceptable at the $93$\%
confidence level. For the best-fit radially-constant model the RMS
residual of the fit is somewhat larger than for the best-fit
radially-varying model, $0.038$ mag vs.~$0.031$ mag, respectively. Its
overall $\chi^2$ is $48.9$.

The viewing angles for the best-fit radially-constant model are $i =
36.5^{\circ} \pm 1.5^{\circ}$ and $\Theta = 121.9^{\circ} \pm
3.3^{\circ}$. The quoted errors are formal 1-$\sigma$ errors
calculated under the assumption of Gaussian random errors in the data
points, and correspond to an increase in the $\chi^2$ of the fit by
$\Delta \chi^2 = 1$. The viewing angles inferred with the
radially-varying models, and the constraints they put on warps and
twists in the LMC disk plane, are discussed in
Section~\ref{s:warptwist}.

\section{Dependence on Photometric Band and Color Selection}
\label{s:difbands}

If the observed spatial variations in the AGB modal magnitude are
indeed due to inclination-induced distance effects, then the
variations should be independent of the photometric band under
study. So it is now useful to consider the LF of stars in all three of
the photometric bands $I$, $J$ and $K_s$, instead of just the $I$
band. The analysis is restricted (for simplicity and for improved
statistics) to a single radial ring, $2.5^{\circ} \leq \rho \leq
6.7^{\circ}$, subdivided in $N=8$ azimuthal sectors (i.e., the sectors
are similar to those shown in Figure~\ref{f:imagegrid}, but are now
not subdivided in four separate rings). Filled points in the top panel
of Figure~\ref{f:banddep} show the results obtained from the $I$-band
LF of the stars with $1.6 \leq I-J \leq 2.0$ (these can be regarded as
an average of the results in the four panels of Figure~\ref{f:fits}).
The open circles and triangles in the same panel are the results
obtained from the $J$ and $K_s$ band LFs. The results in the different
bands agree very well. For comparison, the dashed curve shows the
prediction for the model with $i = 34.7^{\circ}$ and $\Theta =
122.5^{\circ}$ (these values are based on the discussion in
Section~\ref{s:warptwist} below).

The average formal errors in the results are $\langle \Delta \mu_I
\rangle = 0.014$ mag, $\langle \Delta \mu_J \rangle = 0.015$ mag 
and $\langle \Delta \mu_K \rangle = 0.024$ mag. The increase in the
formal error towards larger wavelengths is due to an increase in the
width $\sigma$ (see eq.~[\ref{formalerror}]) of the AGB peak (compare
Figure~\ref{f:CMDs}). To quantify how well the results in the
different bands agree it is useful to define
\begin{equation}
  \chi_{AB}^2 \equiv
     \sum_{l=1}^{L} \sum_{n=1}^{N}
     { {(\mu_{A,ln} - \mu_{B,ln})^2} \over
       {\Delta m_{A,ln}^2 + \Delta m_{B,ln}^2} } .
\label{chisqABdef}
\end{equation}
Here $\mu_{A,ln}$ and $\mu_{B,ln}$ are the results obtained in two
different photometric bands $A$ and $B$, and $\Delta \mu_{A,ln}$ and
$\Delta \mu_{B,ln}$ are the formal errors. The sum is over the $L
\times N$ sectors on the sky, with $L=1$ and $N = 8$ in the present
case. The expectation value ${\cal E}(\chi_{AB}^2)$ of $\chi_{AB}^2$
depends on whether $\mu_{A}$ and $\mu_{B}$ are statistically
correlated. If $\mu_{A}$ and $\mu_{B}$ are statistically independent
estimates of the same underlying quantity, then $\chi_{AB}^2$ should
obey a $\chi^2$ probability distribution with $8$ degrees of freedom,
for which ${\cal E}(\chi_{AB}^2) = 8$. If $\mu_{A}$ and $\mu_{B}$ are
statistically correlated, one expects ${\cal E}(\chi_{AB}^2) <
8$. While individual stellar magnitudes in different bands are based
on data obtained with different detectors, and are therefore
statistically independent, this is not necessarily true for the LFs of
stars selected by color. If stars are selected by $I-J$ color, then
the $I$ and the $J$ band LFs are statistically correlated. If a small
color range is used, then the LFs are almost perfectly correlated (the
LFs then differ only by a constant shift in magnitude). If a very
large color range is used (i.e., almost no color selection), then the
LFs are uncorrelated. The choice $1.6 \leq I-J \leq 2.0$ falls between
these regimes, yielding a partial correlation (Pearson's $r$ between
$0$ and $1$). The results in either band should be less correlated
with the results obtained from the $K_s$-band LF of the same stars,
because the $K_s$ magnitudes were not used in the color
selection. However, even in this case there may be a correlation
because the magnitudes of stars in different bands are intrinsically
correlated (stars in a particular evolutionary phase fall in specific
regions of CMDs and color-color diagrams). For the data in the top
panel of Figure~\ref{f:banddep} one has $\chi_{IJ}^2 = 1.6$,
$\chi_{IK}^2 = 10.7$ and $\chi_{JK}^2 = 9.5$. Given the arguments
mentioned above, this indicates that the results from the different
photometric bands are in statistical agreement.

If the spatial variations in the AGB modal magnitude are due to
inclination-induced distance effects, then it should also not matter
by which color criterion the stars are selected. To verify this the
analysis was repeated using the alternative criterion $1.5 \leq J-K_s
\leq 2.0$ (shown for reference as vertical lines in the bottom left
panel of Figure~\ref{f:CMDs}). The results are shown in the second
panel of Figure~\ref{f:banddep}. The average formal errors in the
three bands are $\langle \Delta \mu_I \rangle = 0.021$ mag, $\langle
\Delta \mu_J \rangle = 0.022$ mag and $\langle \Delta \mu_K \rangle =
0.026$ mag. The results in the different bands are again mutually
consistent, given that $\chi_{IJ}^2 = 3.5$, $\chi_{IK}^2 = 7.4$ and
$\chi_{JK}^2 = 1.7$ (the $J$ and $K_s$-band results are now most
strongly correlated because the stars were selected by $J-K_s$
color). Most importantly, comparison of the top two panels in
Figure~\ref{f:banddep} shows that selection by either $I-J$ or $J-K_s$
color yields results that are statistically indistinguishable.

\section{TRGB Analysis}
\label{s:TRGB}

The inclination of the LMC causes the apparent magnitude of all
features in the CMD to vary with position. So it is not necessary to
restrict the analysis to stars whose nature, physics and stellar
evolutionary properties are well understood. Nonetheless, some caution
is warranted when using AGB stars, since the AGB phase is not
particularly well understood (e.g., Groenewegen \& de Jong 1993) and
individual AGB stars are often variable. So it is useful to repeat the
analysis using a different CMD feature. To this end we have studied
also the tip of the red giant branch. RGB stars are in a different
evolutionary phase than AGB stars, and their properties are governed
by different physics. The mechanism that causes the RGB to have a
well-defined tip is adequately understood (the `Helium Flash'; e.g.,
Chiosi, Bertelli \& Bressan 1992) and the TRGB is commonly used as an
absolute standard candle (e.g., Madore \& Freedman 1995; Salaris \&
Cassisi 1998). A disadvantage of the TRGB is that the formal errors in
the determination of its magnitude are larger than those for the AGB
modal magnitude (cf.~Section~\ref{ss:AGBLF}). This makes it less
useful for quantitative analysis in the present context, but it does
provide an important consistency check.

The analysis was restricted to one radial ring, $2.5^{\circ} \leq \rho
\leq 6.7^{\circ}$, subdivided in $N=8$ azimuthal sections
(in analogy with Section~\ref{s:difbands}). For each section the TRGB
magnitude was determined separately in the $I$, $J$ and $K_s$ bands,
using the algorithm described by Cioni \etal (2000c). The algorithm
does not use any color selection and the Galactic foreground
contribution is subtracted using data for offset fields. The third
panel of Figure~\ref{f:banddep} shows the results for the $I$
band\footnote{The $J$ and $K_s$ band TRGB results, not shown here,
have larger error bars, but are otherwise consistent with the $I$-band
results.}, for which the average formal error is $\langle
\Delta \mu_I \rangle = 0.024$ mag. Comparison to the top
two panels shows that the TRGB results are in good agreement with
those obtained from the AGB modal magnitudes, consistent with the
interpretation that both are the result of inclination-induced
distance variations.

\section{Radial variations of the AGB modal magnitude}
\label{s:varm}

The model fitting procedure of Section~\ref{ss:viewdeterm} includes a
magnitude $m_0$ that must be subtracted from the observed magnitudes
(cf.~eq.~[\ref{magdif}]) to obtain the quantities $\mu$ shown in
Figures~\ref{f:fits} and~\ref{f:banddep}. The quantity $m_0$ is the
apparent magnitude that would have been observed if the stars under
consideration had been positioned at the origin ${\cal O}$ (i.e., the
LMC center). It is similar to an absolute magnitude because it
corresponds to the transformation of an observed apparent magnitude to
a fixed distance (although the actual value of that distance remains
unspecified). The value of $m_0$ is itself not of much interest,
unless one has sufficient theoretical knowledge of the stars under
consideration to be able to predict the {\it absolute} magnitude $M_0$
of the quantity under study (either the AGB modal magnitude or TRGB
magnitude). In that case one can use $m_0$ to determine the distance
modulus of the LMC center. For AGB stars our theoretical understanding
is quite insufficient to use the AGB modal magnitude as an absolute
standard candle; the TRGB has already been used to address that issue
(e.g., Cioni et al.~2000c).

An interesting issue in the present context is to know whether there
are any variations in $m_0$ as function of distance $\rho$ from the
LMC center. This can be addressed for the AGB modal magnitudes (for
the TRGB magnitudes the statistics are insufficient to study this in
much detail). Figure~\ref{f:mzero} shows the inferred radial
dependence of the quantity $m_0 - \langle m_0 \rangle$, where $m_0$ is
the best-fit value for an individual ring, and $\langle m_0 \rangle$
is the average of the $m_0$ values for all rings. Results from the
$I$, $J$ and $K_s$ LFs are shown in separate panels\footnote{Since the
variation in magnitude along a ring is approximately sinusoidal
(cf.~eq.~[\ref{dmagtaylor}]), $m_0$ is approximately equal to the
average of the magnitudes observed for the different azimuthal sectors
along a ring. Consequently, the inferred values for $m_0$ do not
depend in any significant manner on whether or not the viewing angles
$(i,\Theta)$ are allowed to vary in the fit as function of distance
$\rho$ from the LMC center.}. Filled points are for the color
selection criterion $1.6 \leq I-J \leq 2.0$, and open points are for
the criterion $1.5 \leq J-K_s \leq 2.0$. Error bars indicate the
formal 1-$\sigma$ errors under the assumption of Gaussian random
errors in the data points, and correspond to an increase in the
$\chi^2$ of the fit by $\Delta \chi^2 = 1$. The results show that for
the range of radii under study: (a) radial variations in the
distance-corrected AGB modal magnitude are very small, $| m_0 -
\langle m_0 \rangle | \lta 0.03$; and (b) there is a slight but
significant tendency for the distance-corrected AGB modal magnitude to
decrease with increasing distance $\rho$ from the LMC center (${\rm d}
m_0 / {\rm d} \rho \approx 0.01$ mag/degree). The implications of
these findings are discussed below.

\section{Influence of Model Assumptions and Complicating Factors}
\label{s:modassump}

There are significant variations in the apparent magnitude of CMD
features as a function of position in the LMC. These can be
interpreted as the result of variations in distance due to the
inclination of the LMC plane. Such models fit the observed apparent
magnitude variations with a RMS residual that is similar to the formal
errors in the measurements (cf.~Figures~\ref{f:fits}
and~\ref{f:banddep}). This by itself suggests that other effects do
not contribute to the observed apparent magnitude variations at a
level that exceeds the formal measurement errors. Nonetheless, it is
important to address in detail to what extent the analysis may be
influenced by a variety of complicating factors, including: (a)
possible errors in the assumed position of the LMC center; (b)
possible errors in the assumed surface number density distribution;
(c) possible errors in the assumption of a planar geometry; (d)
possible spatial variations in the dust absorption towards or in the
LMC; (e) possible spatial variations in the properties (age,
metallicity, $\ldots$) of the stellar population; and (f) possible
systematic errors in the catalog of stellar magnitudes. Each of these
issues are discussed in turn.

\subsection{Dependence on the Assumed Position of the LMC Center}
\label{ss:LMCcenter}

The LMC has no unique well-defined center. Among other things, the
center of the bar and that of the outer isophotes are offset from each
other by $\sim 0.5^{\circ}$ (Westerlund 1997; Paper II). So the exact
choice of the LMC center in the analysis is to some extent
arbitrary. However, this has no influence on the validity of the
modeling or the accuracy of the results. The equations of
Section~\ref{ss:reldistmag} are equally valid for any choice of the
origin ${\cal O}$, and it is not assumed that the surface number
density $\Sigma(\rho,\Phi)$ would have to be symmetric around ${\cal
O}$. Of course, in practice it does make sense to choose ${\cal O}$
equal to some best estimate for the position of the LMC center, as was
done, if only to ensure that different azimuthal bins have roughly
equal numbers of stars. However, we verified explicitly that a
different choice for the grid center does not yield statistically
different results for the best-fitting viewing angles $(i,\Theta)$.

\subsection{Dependence on the Assumed Surface Number Density Distribution}
\label{ss:numdendep}

The analysis of the viewing angles does not rely on assumptions about
the distribution of stars in the galaxy plane (e.g., circular
symmetry). The surface number density distribution $\Sigma(\rho,\Phi)$
does enter into equation~(\ref{areaintegral}), but only to account for
the relative weighting of different parts of the LMC {\it within} a
sector $A_{ln}$. The sectors $A_{ln}$ adopted for the quantitative
analysis (Figure~\ref{f:imagegrid}) are small enough that the
predicted magnitude variation $\mu(\rho,\Phi;i,\Theta)$ shows only
very modest variations over a sector $A_{ln}$. The model predictions
$\mu_{ln} (i,\Theta)$ are therefore quite insensitive to the
particular form adopted for $\Sigma(\rho,\Phi)$. For the results
presented in the previous sections we adopted an exponential number
density profile (Weinberg \& Nikolaev 2000; Paper II). However,
extensive testing showed that even vastly different assumptions
(including the obviously incorrect assumption of a constant
$\Sigma(\rho,\Phi)$) yield values for $(i,\Theta)$ that agree with
those quoted previously to within the formal errors.

\subsection{The Assumption of a Planar Geometry for the Outer LMC Disk}
\label{ss:planar}

The models that were fit to the data assume that the stars under
consideration lie in a plane. It seems reasonably well established
that the outer geometry of the LMC is indeed planar, as supported by
many lines of evidence. These include the following: (a) the small
vertical scale height indicated by the small line-of-sight velocity
dispersion of Long Period Variables (Bessell, Freeman \& Wood 1986),
star clusters (Freeman, Illingworth \& Oemler 1983; Schommer \etal
1992), planetary nebulae (Meatheringham \etal 1988) and carbon-rich
AGB stars (Alves \& Nelson 2000); (b) the scatter in the
period-luminosity-color relationships for Cepheids (Caldwell \&
Coulson 1986) and Miras (Feast \etal 1989) which would be larger than
observed if the LMC had a significant scale height; (c) the kinematics
of HI (Luks \& Rohlfs 1992; Kim \etal 1998) and other tracers (e.g.,
Schommer \etal 1992) which are well fit by rotating disk models; and
(d) the fact that other Magellanic Irregular galaxies similar to the
LMC, some of which are seen close to edge-on, are known to have small
scale-heights (de Vaucouleurs \& Freeman 1973; McCall 1993). The
actual scale-height of the LMC is probably population dependent, but
the estimates from the line-of-sight velocity dispersion of tracers in
the disk indicate values $\lta 0.5 \kpc$, possibly increasing somewhat
in the outer parts of the disk (Alves \& Nelson 2000). There is no
evidence for a halo component in the LMC comparable to that of our own
Galaxy (Freeman \etal 1983).  At the distance of the LMC ($\sim 51
\kpc$), 1 degree on the sky corresponds to $0.89 \kpc$. The LMC
extends many degrees on the sky, and it is thus reasonable to consider
the LMC geometry planar. It has been a topic of debate whether the LMC
contains secondary populations that do not reside in the main disk
plane (e.g., Luks \& Rohlfs 1992; Zaritsky \& Lin 1997; Zaritsky
\etal 1999; Weinberg \& Nikolaev 2000; Zhao \& Evans 2000), but the
planar geometry of the primary LMC population is not generally called
into question.

The measurements of the apparent magnitude variations along rings are
fit by planar models with an RMS residual of $\sim 0.03$ mag, which is
of the same order of magnitude as the formal errors in the
measurements (cf.~Section~\ref{s:results}). This is an important
result, given that the amplitude of the apparent magnitude variations
are nearly 10 times larger. While this does not necessarily indicate
that the LMC geometry {\it must} be planar, the data presented here
are certainly not in contradiction with this hypothesis. Having said
this, there is no question that the models employed here are in fact
oversimplified because they assume that the stars under consideration
lie in an {\it infinitesimally thin} plane. In reality, the stellar
distribution must extend vertically. So the assumption is made
implicitly that the mean distance to stars along any line of sight can
be approximated as the distance to the equatorial plane. This is
adequate if the vertical extent of the stellar distribution is small
compared to its radial extent, which is supported by the evidence
cited above. For completeness, let us point out that in models with a
considerable thickness one expects smaller magnitude variations along
a ring than in infinitesimally thin models (for a spherical model one
does not expect any magnitude variations). So the inclination values
obtained in this paper would be underestimates if the LMC disk does
have a considerable thickness.

In Section~\ref{s:varm} it was demonstrated that there is a small
decrease with radius of the (distance-corrected) AGB modal magnitude
$m_0$. The decrease is only $\sim 0.04$ mag over the range of $4$
degrees under study, but does appear significant. One possible
explanation for this would be to assume that the LMC is not flat, but
curved towards us like a soup plate. This would cause the stars in the
outermost ring to appear brighter than those closer in. This
explanation does not seem very plausible though, given that such
planar deformations are not generally observed in other disk galaxies
(and also are probably not dynamically stable). Alternatively, one
could get a similar effect if the LMC were not curved, but had both an
increasing scale height with radius and dust in its equatorial
plane. Observations could then possibly be biased towards stars on the
near side of the equatorial plane, and for these stars the average
distance to the observer would decrease as one moves outwards. We have
not constructed detailed models of this type, but maybe such models
could explain the observation that $m_0$ decreases slightly with
radius.

\subsection{Influence of Dust Absorption}
\label{ss:dust}

There is a large amount of dust absorption towards the LMC and
intrinsic to it. The amount of absorption $A_V$ can be $1.0$ mag or
more, although $0.4$ mag may be a more typical value for the cool
stars that are of relevance to the present study (e.g., Zaritsky
1999). In addition, the dust absorption is strongly spatially
variable (e.g., Schwering 1989). So at first glance one might have
thought that any study of the variation in apparent magnitude of CMD
features might have provided more information about dust absorption
than about distance and inclination effects. However, two effects
ameliorate the influence of dust absorption on the analysis. First,
each sector in the analysis is quite large, $\gta 2$ square
degrees. So small scale variations in dust absorption average out, and
the remaining variations are more modest than the variations that are
sometimes seen at small scales (e.g., inside vs.~outside of `dark
clouds'; Hodge 1972). Second, the DENIS survey was performed using
near-IR bands, where the effects of dust absorption are less
pronounced than in the optical. The extinction law given by Glass
(1999) with $R_V = 3.1$ yields for the DENIS passbands that $A_V : A_I
: A_J : A_{K_s} = 1.0 : 0.592 : 0.256 : 0.089$. So even if the average
$A_V$ varied by as much as $0.4$ mag between sectors, then variations
in $A_{K_s}$ would only be $\sim 0.04$ mag. This is of the same order
as the formal errors in the measurements, and more than 5 times less
than the variations expected due to distance effects.

Comparison of the results of the analysis in the different near-IR
photometric bands yields a direct quantitative assessment of the
effects of dust absorption. From Figure~\ref{f:banddep} and the
discussion in Section~\ref{s:difbands} it follows that the azimuthal
variations in the AGB modal magnitude are identical in the $I$, $J$
and $K_s$ bands, to within the formal errors. This would not have been
the case if the observed variations had been due to spatial variations
in dust absorption (in which case the variations would have been
larger in the $I$ band than in the $K_s$ band). Any influence of
azimuthal variations in dust absorption on the observed variations
$\mu$ must therefore be smaller than the formal errorbars $\Delta
\mu$ (i.e., $\lta 0.03$ mag). This is nearly 10 times smaller than the
peak-to-peak amplitude of the observed azimuthal variations. Hence,
possible azimuthal variations in dust absorption do not affect the
analysis at a significant level.

The analysis in Section~\ref{s:varm} provides a constraint on the size
of {\it radial} variations in the average dust absorption. The
distance-corrected AGB modal magnitude $m_0$ decreases by $\sim 0.04$
from the innermost to the outermost ring in the analysis. This could
in principle be due to dust; Figure~\ref{f:mzero} provides a hint that
the variation is smaller in the $K_s$ band than in the $I$ and
$J$-bands, as expect for dust, but the error bars of the $m_0$
measurements are not small enough to test this in detail. More
important in the present context is that the radial variations in
$m_0$ are so small. This indicates that any radial variations in the
azimuthally averaged dust absorption must be quite small as well,
$\Delta A_I \lta 0.04$ mag over a radial range of 4 degrees. So
large-scale variations in dust absorption should have a negligible
influence on the analysis.

\subsection{Influence of Age or Metallicity Effects}
\label{ss:pop}

Stellar magnitudes generally depend on metallicity and age. Therefore,
some of the observed variations in the apparent magnitude of CMD
features could be due to spatial variations in age or metallicity
within the LMC. Spatial variations in age definitely exist, given that
the morphology of the LMC is markedly different for stars in different
age groups (e.g., Cioni \etal 2000b). However, the effect of age
variations on the analysis is ameliorated by the fact that only stars
of a particular type are included in the analysis, either AGB or RGB
stars, and this tends to restrict the analysis to stars of similar
mean age. The existence of a metallicity gradient in the LMC remains a
topic of debate (e.g., Harris 1983; Olszewski \etal 1991; Kontizas,
Kontizas \& Michalitsianos 1993).

An important constraint on the influence of age and metallicity
effects on the analysis is provided by the fact that measurements of
the quantity $\mu$ are available for stars of different types (AGB and
RGB) and in different photometric bands. The magnitude variations are
found to be independent of both the stellar type and the photometric
band, to within the formal errors (cf.~Section~\ref{s:difbands}). This
would not generally be expected if the observed magnitude variations
were due entirely to spatial variations in the age or metallicity of
the population. Such variations often influence stellar magnitudes
differently depending on stellar type and photometric band. So while
this does not prove that stellar population effects are playing no
role at all, there is certainly no indication from the observed
azimuthal variations in $\mu$ that they would.

If at all present, spatial stellar population variations are most
likely to manifest themselves as radial gradients. Galaxies often show
radial color or line-strength gradients which indicate radial changes
in age, metallicity, or both (e.g., Binney \& Merrifield
1998). However, the distance-corrected AGB modal magnitude $m_0$
decreases by only $\sim 0.04$ from the innermost to the outermost ring
in the analysis (cf.~Section~\ref{s:varm}). While this variation could
quite possible be due to radial variations in age or metallicity, its
size is hardly more than the formal errors $\Delta
\mu$ in the measurements. So even if similar stellar-population
induced variations in $\mu$ were to be present in the azimuthal
direction, they would have little influence on the present analysis.

\subsection{Possible Influence of Systematic Errors in the Catalog Magnitudes} 
\label{ss:caterrors}

The DENIS survey strategy uses strips at constant declination, each 12
arcmin wide in RA. The LMC data in the DCMC are made up of 119
different strips that were observed over a period of several years.
The main source of systematic error in the catalog is believed to be
random errors in the zeropoints for the individual strips. For the
data originally discussed by Cioni \etal (2000a) the typical
1-$\sigma$ zeropoint error per strip is in the range $\Delta Z =
0.04$--$0.07$ mag, depending somewhat on the photometric band and on
how the error is estimated. For the analysis in the present paper an
improved calibration of the zeropoints was made, described in
Appendix~\ref{s:AppA}, using the data in the overlap region between
strips. After this new calibration the 1-$\sigma$ zeropoint errors per
strip are only of the order $\Delta Z = 0.01$--$0.02$ mag.

The spatial sectors in the analysis (see Figure~\ref{f:imagegrid}) are
generally $\gta 1$ degree wide in RA, so the data in each sector are
made up of data from $S \gta 5$ different scan strips. The zeropoint
errors between adjacent strips are uncorrelated, since adjacent strips
were usually observed months or years apart. Hence, the average
zeropoint error per sector is expected to be $\sim \Delta Z /
\sqrt{S}$. The systematic zeropoint errors per sector are therefore
expected to be well below the formal errors in the measurements of the
spatial magnitude variations $\mu$, and consequently, such errors
should not affect the analysis at a significant level. Of course,
there could always be some other mysterious systematic error in the
data.  However, the spatial magnitude variations $\mu$ detected in the
catalog are extremely smooth and coherent over the entire area of the
LMC, cf.~Figure~\ref{f:fits}. Since different areas of the LMC were
observed months or years apart, time-ordered more-or-less randomly in
RA, it is hard to think of any type of error that could plausibly
produce such variations.

To further test the possible influence of any possible systematic
errors in the DENIS data, part of the analysis was repeated using the
data from the 2MASS survey. The 2MASS survey obtained data in the $J$,
$H$ and $K_S$ bands, and is in many ways similar to the DENIS
survey. We extracted the 2MASS Point Source Catalog data for the LMC
region of the sky from the Second 2MASS Incremental Data Release. The
analysis was restricted to those stars detected in all three of the
2MASS bands with no special error flags. As in
Section~\ref{s:difbands}, stars were selected with $1.5
\leq J-K_s \leq 2.0$, and these were binned into $N=8$ azimuthal 
sectors\footnote{The 2MASS Second Incremental Data Release does not
yet provide complete coverage of the LMC area (see Paper~II). However,
the missing regions are much smaller than the sectors used in the
analysis, so that the results are not strongly affected by this.}
along one single radial ring, $2.5^{\circ} \leq \rho \leq
6.7^{\circ}$. These data were analyzed similarly as the DENIS data,
yielding the results shown in the bottom panel of
Figure~\ref{f:banddep}. The results are overall in good agreement with
those obtained with the same color selection criterion from the DENIS
data (second panel of Figure~\ref{f:banddep}), and indeed with all of
the results obtained from the DENIS data. Hence, any systematic errors
that may be present in any of the two surveys do not have a
significant impact on the main results of our study.

\subsection{Influence of Foreground and Background Sources}
\label{ss:foreandback}

The DENIS and 2MASS catalogs in the direction of the LMC are of course
contaminated by foreground and background sources. The main foreground
contamination comes from Galactic disk stars and the main background
contamination comes from galaxies behind the LMC. However, these do
not affect the analysis at a noticeable level. The AGB star analysis
is restricted to stars with red colors, which efficiently eliminates
Galactic Disk stars (Nikolaev \& Weinberg 2000). Background galaxies
tend to be fainter than the LMC AGB stars (Nikolaev \& Weinberg 2000),
so while they do contribute to the faint end of LFs such as those in
Figure~\ref{f:histograms}, they do not affect the Gaussian fits to the
LF peak (described in Section~\ref{ss:LFanalysis}). Foreground and
background sources also should not affect the TRGB analysis, given
that their LF is smooth and continuous near the RGB tip. This was
addressed explicitly in Section~A.3.3 of Cioni \etal (2000c).  Note
that our procedures for foreground and background elimination are
quite different for the AGB and TRGB analyses. The excellent agreement
between the results from these analyses (cf.~Figure~\ref{f:banddep})
therefore provides further evidence that foreground and background
contamination do not introduce systematic errors.

\section{Warps and Twists of the LMC plane} 
\label{s:warptwist}

Having established that the observed spatial variations in the
apparent magnitude of CMD features can only be plausibly interpreted
as the result of inclination-induced distance variations, it is now
appropriate to consider the issue of possible warps and twists in the
LMC plane. The spatial grid on the sky shown in
Figure~\ref{f:imagegrid} uses four radial rings that span the range
$2.5^{\circ} \leq \rho \leq 6.7^{\circ}$. Figure~\ref{f:fits} showed
the results for these rings obtained from an analysis of the $I$-band
LF of AGB stars selected by $I-J$ color, and Section~\ref{s:results}
discussed model fits with constant viewing angles $(i,\Theta)$. One
may alternatively fit models in which $(i,\Theta)$ are allowed to vary
as function of the distance $\rho$ from the LMC center. Solid curves
in Figure~\ref{f:fits} show the best-fit results for these
`radially-varying' models. Filled points in the top panels of
Figure~\ref{f:viewang} show the inferred $(i,\Theta)$ as function of
angular distance $\rho$. For comparison, open points show the viewing
angles for the best-fit radially-constant model, for which $i =
36.5^{\circ} \pm 1.5^{\circ}$ and $\Theta = 121.9^{\circ} \pm
3.3^{\circ}$ (note that the radially constant model fits the $\mu$
data with a poorer $\chi^2$ than the radially-varying model,
cf.~Section~\ref{s:results}).

There is a trend for both $i$ and $\Theta$ to decrease with increasing
distance $\rho$ from the LMC center. As a test of the robustness of
this result, the middle panels of Figure~\ref{f:viewang} show the
results of a similar analysis for AGB stars selected from the 2MASS
survey with the color selection criterion $1.5 \leq J-K_s \leq 2.0$.
The viewing angles were determined by fitting simultaneously the
$\mu$-values inferred from the $J$, the $H$ and the $K_s$ band LFs of
these stars. The viewing angles for the best-fit radially-constant
model are $i = 34.5^{\circ} \pm 1.4^{\circ}$ and $\Theta =
123.5^{\circ} \pm 3.0^{\circ}$, consistent with the values quoted
above for the analysis of the $I$-band LF of AGB stars selected by
$I-J$ color from the DENIS survey. However, the results for the
radially-varying models are somewhat different for the two
analyses. The results for the variation of $\Theta$ with $\rho$ are
consistent given the error bars, but while the top panels of
Figure~\ref{f:viewang} suggest a decline of $\Theta$ with $\rho$, no
evidence for this is seen in the middle panels. For the inclination,
the results are mutually consistent only for the first two radial
rings ($\rho \leq 4.4^{\circ}$); the results for the outer two rings
differ by $\sim 3\sigma$.

In an attempt to reduce as much as possible all sources of error, one
final overall combined fit was performed to the variations in $\mu$
inferred from separate analyses of the following data: (a) $I$, $J$
and $K_s$ band DENIS data of stars selected to have $1.6 \leq I-J \leq
2.0$; (b) $I$, $J$ and $K_s$ band DENIS data of stars selected to have
$1.5 \leq J-K_s \leq 2.0$; and (c) $J$, $H$ and $K_s$ band 2MASS data
of stars selected to have $1.5 \leq J-K_s \leq 2.0$. The bottom panels
of Figure~\ref{f:viewang} show the results. The viewing angles for the
best-fit radially-constant model are $i = 34.7^{\circ} \pm
0.7^{\circ}$ and $\Theta = 122.5^{\circ} \pm 1.6^{\circ}$. The errors
on these numbers reflect only the propagation of random errors into
the inferred viewing angles. They should therefore be interpreted with
some scepticism. The differences between the results inferred from the
DENIS and 2MASS data (top and middle panels of Figure~\ref{f:viewang})
indicate that there are probably small systematic errors in the
analysis as well, which may not disappear by averaging. In addition,
there is some evidence for real variations in the viewing angles with
$\rho$, as further discussed below. A more conservative estimate of
the errors is therefore obtained by calculating the dispersion in the
individual measurements of $i$ and $\Theta$ obtained from the
different radial rings in different datasets. This yields
$6.2^{\circ}$ and $8.3^{\circ}$, respectively. So as the final results
of this paper we adopt $i = 34.7^{\circ} \pm 6.2^{\circ}$ and $\Theta
= 122.5^{\circ} \pm 8.3^{\circ}$.

The results of the best-fit radially-varying model (bottom panels of
Figure~\ref{f:viewang}) suggest that there is an abrupt decrease in
inclination from $i \approx 40^{\circ}$ for $\rho < 4.4^{\circ}$ to $i
\approx 31^{\circ}$ for $\rho > 4.4^{\circ}$, as well as a gradual
decrease in the position angle of the line of nodes from $\Theta
\approx 125^{\circ}$ for $\rho \approx 3^{\circ}$ to 
$\Theta \approx 115^{\circ}$ for $\rho \approx 6^{\circ}$.  While
systematic errors may play some role in this, the variations could
well be real. One natural interpretation would be that the LMC disk
plane is warped. This would, by definition, cause the inclination to
vary with $\rho$. Since the line of nodes of the warp is physically
unrelated to the line of nodes of the galaxy on the sky, a warp
typically (but not necessarily) also induces a twist in the position
angle of the line of nodes $\Theta$ with radius. On the other hand,
warping and twisting of the LMC disk plane is not the only viable
explanation for the inferred radial variations in the viewing
angles. It was pointed out in Section~\ref{ss:planar} that our models
would underestimate the inclination of the LMC if its disk had a very
considerable vertical thickness. So the observed radial decrease in
$i$ could in principle be due to a radial increase in the vertical
scale height of the LMC disk. While it has been found that such
behavior is not typical for spiral galaxies (e.g., van der Kruit \&
Searle 1981), it is unknown whether this result holds for later type
galaxies as well. In fact, Alves \& Nelson (2000) have suggested that
the vertical thickness of the LMC does indeed increase with radius, based
on the fact that the line-of-sight velocity dispersion of carbon stars
in the LMC disk does not fall with radius. While this may be able to
qualitatively explain the (apparent) radial decrease in inclination,
detailed modeling would be required for a more quantitative assessment
of this hypothesis.

\section{Comparison to Previous Estimates of the LMC Viewing Angles}
\label{s:comparison}

A comprehensive summary of previous work on the LMC viewing angles is
provided in Table~3.5 of the book by Westerlund (1997). The generally
quoted consensus that has emerged from these studies is that the
inclination angle $i$ is somewhere in the range
$25^{\circ}$--$45^{\circ}$, and that the position angle of the
line-of-nodes is somewhere in the range
$140^{\circ}$--$190^{\circ}$. The inclination angle inferred here, $i
= 34.7^{\circ} \pm 6.2^{\circ}$, is comfortably within the range of
previously quoted values. However, the position angle of the line of
nodes, $\Theta = 122.5^{\circ} \pm 8.3^{\circ}$, is very different
from the values that have generally been quoted. To achieve an
understanding of this discrepancy, it is useful to discuss the various
methods that have previously been used to estimate the LMC viewing
angles.

\subsection{The Photometric Circular Disk Method}
\label{ss:circphot}

The method that has been used most often to estimate the LMC viewing
angles is what we will refer to as the `photometric circular disk
method'. It assumes that the intrinsic shape of the LMC disk (at large
radii) is circular. If this is true, then the major axis of the
projected elliptical shape on the sky coincides with the line of
nodes. Since the major axis position angle $\Theta_{\rm maj}$ is
directly observable, this yields a simple estimate for the position
angle $\Theta$ of the line of nodes; the inclination can be estimated
as $i = \arccos (1-\epsilon)$, where $\epsilon$ is the apparent
ellipticity.  This method has been applied to many of the different
tracers that are available in the LMC disk, including: (a) optical
isophotes of starlight (de Vaucouleurs \& Freeman 1973; Bothun \&
Thompson 1988; Schmidt-Kaler \& Gochermann 1992); (b) contours of the
number density of stars detected in the near-IR (Weinberg \& Nikolaev
2000), of stellar clusters (Lynga \& Westerlund 1963;
Kontizas \etal 1990) or of HII
regions, supergiants, or planetary nebulae (Feitzinger, Isserstedt \&
Schmidt-Kaler 1977); and (c) the brightness contours of HI emission
(McGee \& Milton 1966; Kim \etal 1998) or non-thermal radio emission
(Alvarez, Aparici \& May 1987). Results obtained from these studies
have generally fallen in the range $i = 20^{\circ}$--$45^{\circ}$ and
$\Theta_{\rm maj} = 160^{\circ}$--$190^{\circ}$. The variation in the
results from different authors may be due in part to differences in
the distance of the tracers to the LMC center, given that the LMC has
considerable radial gradients in both the ellipticity and the major
axis position angle of its contours (see Paper~II).

The main disadvantage of the photometric circular disk method is that
it makes the ad hoc assumption that the LMC disk is circular.  While
this seems reasonable at first glance, there really is no a priori
reason why galaxy disks should be circular. It is possible to
construct self-consistent dynamical models for elliptical disks (e.g.,
Teuben 1987), and it is known that bars and other planar
non-axisymmetric structures are common in disk galaxies. The dark
matter halos predicted by cosmological simulations are generally
triaxial (e.g., Dubinski \& Carlberg 1991), and the gravitational
potential in the equatorial plane of such halos does not have circular
symmetry. So disks are expected to be elongated, and this has been
confirmed for those galaxies that have been studied in sufficient
detail to address this issue (e.g., Schoenmakers, Franx \& de Zeeuw
1997). For the LMC there is the additional argument that it is both
moving in the tidal field of the Galaxy and interacting with the SMC,
both of which may have distorted its shape.

The average result obtained here from the apparent magnitude
variations along rings, $\Theta = 122.5 \pm 8.3^{\circ}$, is quite
inconsistent with the values $\Theta_{\rm maj}$ that have been
obtained from the photometric circular disk method. In other words,
the line of nodes of the LMC is {\it not} coincident with the major
axis of the distribution of disk tracers on the sky. This provides
important new information on the structure of the LMC: it implies that
{\it the LMC disk is not intrinsically circular}. Paper~II will
explore this conclusion through a detailed analysis of the LMC shape
and structure.

\subsection{The Kinematic Circular Disk Method}
\label{ss:circkin}

In the `kinematic circular disk method' the viewing geometry of the
LMC is estimated by interpreting the observed line-of-sight velocities
of tracers in the disk under the assumption of intrinsically circular
orbits. This method has been applied to various tracers, including HI
(Rohlfs \etal 1984; Luks \& Rohlfs 1992; Kim \etal 1998), star
clusters (Freeman \etal 1983; Schommer \etal 1992), planetary nebulae
(Meatheringham \etal 1988), HII regions and supergiants (Feitzinger
\etal 1977), and carbon-rich AGB stars (Kunkel \etal 1997; Graff \etal 
2000; Alves \& Nelson 2000). Analysis yields the position angle
$\Theta_{\rm max}$ of the `kinematic line of nodes', defined as the
line of maximum velocity gradient. For a circular model this coincides
with the true line of nodes (the intersection of the plane of the
galaxy and the plane of the sky). The inclination is not generally
well constrained by the observed velocity field, because the $\sin i$
component in the observed velocities can be roughly cancelled by
modifications in the (unknown) intrinsic rotation curve amplitude (in
the case of solid body rotation this degeneracy is complete; see
Schoenmakers \etal 1997). The general procedure in the analysis has
therefore often been to fix $i$ a priori, usually to a value estimated
from the photometric circular disk method.

After correction for the transverse motion of the LMC (e.g., Kroupa \&
Bastian 1997; Alves \& Nelson 2000), the kinematic line of nodes for
the available tracers has generally been found to be in the range
$\Theta_{\rm max} = 140^{\circ}$--$190^{\circ}$. This is more-or-less
consistent with the values inferred from the photometric circular disk
method, and is inconsistent with the average result $\Theta = 122.5
\pm 8.3^{\circ}$ inferred here from the apparent magnitude variations
along rings. However, if the LMC disk is elliptical instead of
circular, one expects a misalignment between $\Theta$ and $\Theta_{\rm
max}$ (e.g., Franx, van Gorkom \& de Zeeuw 1994; Schoenmakers \etal
1997). The HI velocity field and discussion presented by Kim \etal
(1998) actually support this interpretation, by showing that the
kinematic principle axes are not perpendicular to each other, and that
$\Theta_{\rm max}$ twists by $\sim 20^{\circ}$ from small to large
radii. In Paper~II we discuss the kinematics of the LMC in detail, and
address how the observed kinematics can be interpreted in the context
of the line of nodes position angle inferred here.

\subsection{The Relative Distance Variation Method applied to Cepheids}
\label{ss:cepheids}

The most direct (and hence most accurate) way to determine the viewing
angles of the LMC is with the `relative distance variation method',
for which the theoretical formalism was presented in
Section~\ref{s:theory}. This method only uses geometry, with no
assumptions about either the distribution or kinematics of tracers in
the LMC plane. While the method was applied here to late-type stars,
its primary use has so far been to analyze data on Cepheids. Several
detailed studies on this topic were published in the 1980s. Caldwell
\& Coulson (1986) analyzed optical data for 73 Cepheids and obtained
$i = 29^{\circ} \pm 6^{\circ}$ and $\Theta = 142^{\circ} \pm
8^{\circ}$. They used `statistical reddenings' for the majority of
their stars. Laney \& Stobie (1986) and Welch
\etal (1987) both used near-IR data with individually determined reddenings, 
but had correspondingly smaller samples. Laney \& Stobie obtained $i =
45^{\circ} \pm 7^{\circ}$ and $\Theta = 145^{\circ} \pm 17^{\circ}$
from 14 Cepheids, and Welch \etal obtained $i = 37^{\circ} \pm
16^{\circ}$ and $\Theta = 167^{\circ} \pm 42^{\circ}$ from 23
Cepheids.

These Cepheid studies do not provide very strong constraints on the
LMC viewing geometry, given the small sample sizes and correspondingly
large statistical uncertainties. Nonetheless, it is interesting to
note that for the studies with smallest error bars (Caldwell \&
Coulson 1986; Laney \& Stobie 1986) the estimates for $\Theta$ of $142
\pm 8^{\circ}$ and $145^{\circ} \pm 17^{\circ}$ are significantly
lower than most values that have been obtained from the photometric
and kinematic circular disk methods. This is qualitatively similar to
the main result obtained here, and provides independent evidence that
the LMC is not circular. The average value $\Theta = 122.5^{\circ} \pm
8.3^{\circ}$ obtained here differs from the Caldwell \& Coulson value
at the 2-$\sigma$ level, and the best-fit inclinations differ at the
$1$-$\sigma$ level. However, these differences should not necessarily
be viewed as significant; the statistical errors of Caldwell \&
Coulson may well be underestimates of the true errors, given the use
of statistical reddenings instead of individually determined
reddenings for the majority of their Cepheids. More importantly, the
accuracy of the results presented here is superior to those obtained
from Cepheid studies due to the much larger number of available stars
and the detailed assesment of systematic effects.

Recent LMC surveys that search for microlensing events such as MACHO
(Alcock \etal 1999), EROS (Beaulieu \etal 1995) and OGLE (Udalski
\etal 1999) have found of order a thousand new Cepheids. These new 
samples should allow improvement over the older Cepheid-based
studies. On the other hand, these samples were obtained from
observations that focused on the bar of the LMC. So they cover only
small radii $\rho$, where any distance-induced magnitude variations
are expected to be small (cf.~eq.~[\ref{dmagtaylor}]). Also, the
Cepheids tend to fall, by observational construction, along the bar
(see e.g.~Figure~1 of Udalski \etal 1999 and Figure~1 of Alcock \etal
2000b), which is a linear structure on the sky. This makes it
difficult to study the variation of the Cepheid magnitudes as function
of the azimuthal angle $\Phi$, so that the position angle $\Theta$ of
the line of nodes can probably only be poorly constrained.

The only study so far that uses the new Cepheid samples is that of
Groenewegen (2000) of OGLE database (Udalski \etal 1999). However, his
analysis is not truly based on the relative distance variation
method. The position angle $\Theta$ of the line of nodes is not
determined from the relative distance variations of the Cepheids, but
is instead fixed to be perpendicular to the major axis position angle
$\Theta_{\rm maj}$ of the Cepheid distribution on the sky, i.e.,
$90^{\circ}$ plus the major axis position angle of the bar, which
yields\footnote{We subtracted $90^{\circ}$ from the value quoted by
Groenewegen (2000) to obtain the value appropriate for the coordinate
systems defined in the present paper.} $\Theta = 206^{\circ} \pm
0.5^{\circ}$. The determination of the inclination angle in
Groenewegen's analysis is then similar to that in the relative
distance variation method; i.e., $i$ is determined from the magnitude
variations of the Cepheids along a line perpendicular to the adopted
line of nodes (i.e., along the bar). This yields $i = 18^{\circ} \pm
3^{\circ}$. A direct comparison of the viewing angles inferred here to
those of Groenwegen may not be meaningful. The Cepheids in his study
all fall in the inner parts of the LMC, a region that has been
specifically excluded from the study in the present paper. The
difference between his results and those presented here could
therefore in principle be ascribed to real variations in the viewing
angles as a function of radius. However, we believe that such a
drastic interpretation is not called for, given that Groenewegen's
assumed value for $\Theta$ is completely arbitrary. There is no good
reason why the angle adopted by him should bear any physical relation
to the actual position angle of the line nodes. It seems reasonable to
attribute the fact that Groenewegen's results for $(i,\Theta)$ are
inconsistent with those derived here (and with the majority of all
other values quoted in the literature) to this ad hoc assumption
underlying his analysis.

\subsection{The Relative Distance Variation Method applied to 
2MASS AGB Star Data}
\label{ss:AGB2MASS}

The approach of using AGB modal magnitudes to study distance
variations in the LMC was used and advocated previously by Weinberg \&
Nikolaev (2000). They selected AGB stars from the 2MASS survey data
based on the $J-K_S$ color, and focused on obtaining an AGB LF peak
that is narrow even in the $K_S$ band. The $(J-K_s,K_s)$ CMD in the
bottom right panel of Figure~\ref{f:CMDs} shows that this rules out
the use of the O-rich AGB stars, which have a spread in $K_S$
magnitude of more than a full magnitude. The C-rich AGB stars also
spread over a full magnitude in $K_s$, but their $K_s$ magnitudes
correlate strongly with $J-K_s$ color. So one does obtain a reasonably
narrow LF peak if one restricts the analysis to a small range in
$J-K_s$ color. Weinberg \& Nikolaev adopted stars with $1.6
\leq J-K_S \leq 1.7$ for their study. The DENIS Survey has data in the
$I$-band, which is not available with 2MASS. This gives the option to
select stars by $I-J$ color for the present study, which we found to
yield superior statistics. Nonetheless, Figure~\ref{f:banddep} shows
that selecting stars by $J-K_s$ color does not yield appreciably
different results. If one uses a large range of $J-K_s$ colors, the
error bars increase only by a factor $\sim 1.5$ as compared to
selection by $I-J$ color. However, Weinberg \& Nikolaev adopted a
range of $J-K_s$ colors that is 5 times smaller than what was used
here for Figure~\ref{f:banddep}. As a consequence, they ended up with
a much smaller sample of stars for their analysis. Selection of stars
with the criterion $1.6 \leq I-J \leq 2.0$ from the DENIS catalog
yields $\sim 10$ times more stars in the AGB peak than selection with
the criterion $1.6 \leq J-K_S \leq 1.7$, with no significant
difference in the width $\sigma$ of the AGB peak. Therefore, the
results of the present analysis are considerably more accurate than
those of Weinberg \& Nikolaev.

Weinberg \& Nikolaev did not make a very fine subdivision of the LMC
area for their study, presumably forced by the smaller statistics.
They studied the LF of AGB stars for 9 fields: one field on the LMC
center, and 2 fields in each of the directions North, East, South and
West, respectively. From their analysis they found a considerable
$\mu$-gradient in the E-W direction, but no gradient in the N-S
direction. From this they concluded that the position angle of the
line of nodes is consistent with the value $\Theta_{\rm maj} \approx
170^{\circ}$ that they derived from the major axis position angle of
the stellar number density contours. This is in direct contradiction
with the results obtained here, both from DENIS data and from the same
2MASS data. The analysis presented here yields significant gradients
in both the E-W and N-S directions (and it was verified that this is
also the case if the analysis is restricted to exactly the same LMC
fields as studied by Weinberg \& Nikolaev). This disagreement is most
likely due to the limited statistics of the Weinberg \& Nikolaev
analysis. The values of $\mu$ increase with radius $\rho$
(cf.~Figure~\ref{f:maxminvar}), and most of the weight in Weinberg \&
Nikolaev's results therefore comes from their 4 outermost fields at
$\sim 5^{\circ}$ from the LMC center. In these fields the available
numbers of stars in their analysis are 31, 29, 18 and 39, in the
directions N, E, S, W, respectively (cf.~their figures~7
and~8). Weinberg \& Nikolaev do not discuss at all how they derive
either $\mu$ or formal errorbars $\Delta \mu$ from the stellar
magnitudes, but they do show errorbars $\Delta \mu \approx 0.03$ mag
for these fields in their figure~9. However, the results of the
Monte-Carlo simulations discussed in Section~\ref{ss:LFanalysis}
suggest that it is not possible to obtain estimates for $\mu$ that are
as accurate as this, when only so few stars are available. A visual
inspection of the LF histograms that Weinberg \& Nikolaev present for
these fields appears to confirm this. The histograms are quite
unsmooth and skewed, and their peaks are in many cases significantly
($\sim 0.2$ mag) offset from the peaks of the smooth-kernel estimates
used to estimate $\mu$. So while the Weinberg \& Nikolaev suggestion
to use near-IR AGB modal magnitudes for relative distance measurements
was very important, their inferred $(i,\Theta)$ values are probably
not accurate. Their results are certainly not supported by an analysis
of similar detail as that presented here.

\section{Distances to some well-studied objects}
\label{s:indpos}

The distance of the LMC has remained a controversial subject (e.g.,
Mould \etal 2000; Udalski 2000). Some methods for the determination of
the LMC distance are based on observations of a single object, such as
SN 1987A (Panagia \etal 1991; McCall 1993), or the eclipsing binaries
HV 982 (e.g., Fitzpatrick \etal 2001) and HV 2274 (Nelson \etal 2000).
In such cases the results must be corrected for the relative distance
$D/D_0$ of the object with respect to the LMC center. If ones assumes
that these objects reside in the plane defined by the outer LMC disk,
then $m-m_0$ can be calculated from equations~(\ref{Dsimpler})
and~(\ref{magdif}). Since the viewing angles derived here differ
considerably from earlier estimates, it is useful to recalculate
$m-m_0$ for these objects. With the average values of $(i,\Theta)$
derived in Section~\ref{s:warptwist} this yields $m-m_0$(SN
1987A)$=-0.013$, $m-m_0$(HV 982)$=-0.009$ and $m-m_0$(HV
2274)$=0.015$. These corrections are all small, primarily because the
objects do not lie far from the LMC center. The corrections that have
previously been applied to distance determinations from these objects
have been similarly small. Hence, the new values for the viewing
angles $(i,\Theta)$ derived here do not have much impact on the
previous determinations of the LMC distance scale.

\section{Conclusions}
\label{s:conc}

We have presented a detailed study of the LMC viewing angles, using
the stars detected in the DENIS survey at distances $\rho$ between
$2.5^{\circ}$ and $6.7^{\circ}$ from the LMC center. For an inclined
disk one expects a sinusoidal variation in the brightness of tracers
as function of position angle along a circle. We detect such
brightness variations at high confidence from an analysis of the
apparent magnitude of features in the near-IR CMDs. The peak-to-peak
amplitude of the variations is $\sim 0.25$ mag. The same variations
are detected for AGB stars (using the mode of their LF) and for RGB
stars (using the tip magnitude derived from their LF). They are seen
consistently in all three of the DENIS bands, $I$, $J$ and $K_s$, and
are seen in the data from the 2MASS survey as well. Any radial
variations in the characteristic magnitudes of the AGB stars are small
($\lta 0.04$ mag). The results of these analyses and our discussions
of these facts argue overwhelmingly that the observed brightness
variations are due to distance variations. Any complicating effects,
such as possible spatial variations in dust absorption or the
age/metallicity of the stellar population, do not appear to cause
brightness variations at a level that exceeds the formal measurement
errors ($\sim 0.03$ mag).

The observed spatial brightness variations are well fit by a geometric
model of an inclined plane. In the best-fit model the average
inclination angle is $i = 34.7^{\circ} \pm 6.2^{\circ}$ and the
average line-of-nodes position angle is $\Theta = 122.5^{\circ} \pm
8.3^{\circ}$. The quoted errors are conservative estimates that take
into account the possible influence of systematic errors. The formal
errors on the viewing angles are much smaller, $0.7^{\circ}$ and
$1.6^{\circ}$, respectively. There is tentative evidence for
variations in the viewing angles with distance $\rho$ from the LMC
center, in the sense of there being an abrupt decrease in inclination
from $i \approx 40^{\circ}$ for $\rho < 4.4^{\circ}$ to $i
\approx 31^{\circ}$ for $\rho > 4.4^{\circ}$, as well as a gradual
decrease in the position angle of the line of nodes from $\Theta
\approx 125^{\circ}$ for $\rho \approx 3^{\circ}$ to 
$\Theta \approx 115^{\circ}$ for $\rho \approx 6^{\circ}$. This may
indicate that the LMC disk plane is warped.

The large majority of all previous studies of the LMC viewing geometry
have measured either the major axis position angle $\Theta_{\rm maj}$
of the spatial distribution of tracers on the sky, or the position
angle $\Theta_{\rm max}$ of the kinematic major axis (the line of
maximum velocity gradient). These studies have generally yielded
values between $140^{\circ}$ and $190^{\circ}$. For a circular disk
one always has that $\Theta_{\rm maj} = \Theta_{\rm max} = \Theta$,
and the observationally determined values for $\Theta_{\rm maj}$ and
$\Theta_{\rm max}$ have therefore generally been quoted in the
literature as the position angle of the line of nodes for the LMC.
Previous studies of spatial brightness variations of Cepheids or AGB
stars have either not had sufficient statistics or have not been
sufficiently detailed to address the LMC viewing angles with the same
accuracy as obtained here. Consequently, these studies have generally
been interpreted as being broadly consistent with the results obtained
from the distribution and kinematics of LMC tracers under the
assumption of circular symmetry.

Our study of the LMC geometry from the distance/brightness variations
of tracers in the disk is the most accurate study of its kind to date,
not only in terms of formal errors, but quite importantly, also in
terms of its control of possible systematic errors. This allows us to
test for the first time with reasonable accuracy to what extent the
assumption of circular symmetry is justified for the LMC disk. We find
that $\Theta$ differs considerably from both $\Theta_{\rm maj}$ and
$\Theta_{\rm max}$. This indicates that the intrinsic shape of the LMC
disk is not circular, but elliptical. The inclination angle inferred
here is broadly consistent with the values that have generally been
quoted in the literature. However, most previous determinations were
based on the incorrect assumption of circular symmetry, so this is to
some extent a coincidence. In Paper~II of this series we explore in
detail the implications of the newly derived viewing angles through a
detailed study of the shape and structure of the LMC.


\acknowledgments

We are grateful to all members of the DENIS consortium for their role
in the collection of the DENIS data, and in particular to Cecile Loup
for assisting with the most recent updates to the DCMC. Part of the
analysis made use of data products from the Two Micron All Sky Survey,
which is a joint project of the University of Massachusetts and the
Infrared Processing and Analysis Center/California Institute of
Technology, funded by the National Aeronautics and Space
Administration and the National Science Foundation. We thank Harm
Habing for useful discussions. The anonymous referee provided useful
comments that helped improve the presentation of the paper.

\clearpage


\appendix

\section{Photometric calibration and accuracy}
\label{s:AppA}

The construction of the DENIS Catalog towards the Magellanic Clouds
(DCMC) was discussed in Cioni \etal (2000a). In this Appendix we
discuss the accuracy of the catalog, which is of particular importance
for the research presented here. We also present a new calibration
scheme that was used to further improve the photometric accuracy.

There are two kinds of errors in the derived stellar magnitudes:
formal (i.e., random noise-related) errors and systematic errors. The
formal errors depend primarily on the exposure time adopted for the
survey, and were discussed in Cioni \etal (2000c). These errors are of
course largest for faint stars. However, the results in the present
paper depend primarily on the properties of relatively bright AGB
stars with typical magnitudes of $I \approx 14.1$, $J \approx 12.3$
and $K_s \approx 10.9$. At these magnitudes the average formal errors
are very small: $\Delta I = 0.01$, $\Delta J = 0.01$ and $\Delta K_s =
0.03$. Furthermore, the analysis in the present paper depends on the
average magnitudes of groups of stars in the CMD. Averaging reduces
the influence of formal errors on the end-result by a factor
$\sqrt{N}$ compared to the errors for individual stars, where $N$ is
the number of stars under consideration. Formal errors therefore have
a negligible influence on the analysis in this paper.

To assess systematic errors it is necessary to discuss first the DENIS
survey strategy (Epchtein \etal 1997). The strategy has been to divide
the (southern) sky in three declination zones, each of which was
subdivided into strips at different values of right ascension
(RA). Each strip is 12 arcmin wide in RA (the field of view of an
individual DENIS image), and $30^{\circ}$ long in declination. In the
southernmost zone, which contains the Magellanic Clouds, the distance
in RA between the centers of adjacent strips is 1m$\>$20s. The LMC
data in the DCMC were constructed from the analysis of the 119 strips
with RA between 4h$\>$06m and 6h$\>$47m. Each strip is composed of 180
images at different declinations; adjacent images overlap by 2 arcmin
in declination. The 180 images are observed in a single night, and are
reduced as a unit (Cioni \etal 2000a). On average, 8 strips are
observed per night, but only one or two of these cover the Magellanic
Clouds. Standard star observations are used to determine one
photometric zeropoint for each night, in each of the three broad-bands
$I$, $J$ and $K_S$. The zeropoint, $Z$, is used to transform a fully
calibrated (bias-subtracted, flat-fielded, atmospheric extinction
corrected, etc.)  count rate per second, $C$, to a magnitude, $m$,
using an equation of the form $m = Z - 2.5 \log C$.

The zeropoint $Z$ characterizes the telescope/instrument system, and
should therefore be fairly constant. This is indeed the case.  The
average and RMS of the zeropoints for the strips that make up the DCMC
(observed in the four year period from December 1995 to November 1999)
are $Z_I = 23.41 \pm 0.11$, $Z_J = 21.11 \pm 0.15$, and $Z_K = 19.13
\pm 0.18$. Close inspection of the zeropoints does show 
trends with time, both long-term and short term, some of which can be
directly related to known changes in the telescope/instrument system
(changes in gain, instrument cleanings, etc.). The fact that the
zeropoint is calibrated separately for each night ensures that such
issues have no systematic effect on the catalog.

While it is essential that the zeropoint is calibrated separately for
each night, it is important to realize that each inferred zeropoint is
only known with a certain formal error $\Delta Z$. This formal error
can be estimated from the observations (using the fact that for each
night multiple standard star observations are available, each of which
provides a separate estimate of the zeropoint). For the strips that
make up the LMC data in the DCMC, the average zeropoint errors are
$\langle \Delta Z_I \rangle = 0.04$, $\langle \Delta Z_J \rangle =
0.05$, and $\langle \Delta Z_K \rangle = 0.05$. These zeropoint errors
are the dominant source of photometric error in the catalog. An error
in the zeropoint determination for a given strip causes {\it all} the
stellar magnitudes in that strip to be in error by that amount. So in
this sense the errors are systematic. However, for the catalog as a
whole the zeropoint errors should average to zero. Hence, distance
determinations using the whole catalog (e.g., from the magnitude of
the TRGB; see Cioni \etal 2000c) should not contain a systematic bias.
 
Independent information on the photometric accuracy of the catalog can
be obtained from the overlap region between adjacent
strips. Observations for adjacent strips overlap in RA by 2 or more
arcmin (depending on declination), so there are many stars in each
overlap region for which a magnitude determination is available from
two different observations. Adjacent strips were not generally
observed closely separated in time; more often than not, observations
were obtained months or years apart. Hence, for most realistic sources
of error, the observations in adjacent strips can be considered to be
independent.

Let $l$ and $n$ be the index numbers of two adjacent strips, and let
$m_l$ and $m_n$ denote stellar magnitudes determined from the data for
the respective strips. Let $D_{ln}$ be the average magnitude
difference for all the stars in the overlap region: $D_{ln} \equiv
\langle m_l - m_n \rangle$. A non-zero value of $D_{ln}$ indicates 
that there is a shift in the magnitude scale between the strips. For
each set of adjacent strips we estimated $D_{ln}$ from the data (see
also Figure~1 of Cioni \etal 2000a), together with a formal error
$\Delta D_{ln}$. The RMS values of the $D_{ln}$ thus obtained are
${\rm RMS}(D_{ln})_I = 0.07$, ${\rm RMS}(D_{ln})_J = 0.10$, and ${\rm
RMS}(D_{ln})_{K_s} = 0.10$. In an average sense, these numbers provide
a direct estimate of the zeropoint error per strip, through the
formula $dZ \equiv {\rm RMS}(D_{ln}) / \sqrt{2}$ (the $\sqrt{2}$
arises because $D_{ln}$ is a difference in magnitude between two
strips that both have an independent zeropoint error, and because
errors add in quadrature). This yields $dZ_I = 0.05$, $dZ_J = 0.07$
and $dZ_{K_s} = 0.07$. These numbers are very similar to the average
formal errors $\langle \Delta Z \rangle$ inferred from standard star
observations for each strip (see above), which provides a successful
consistency check.

For the analysis in the present paper, which studies magnitude
differences between stars in nearby areas of the sky, zeropoint errors
are more of a problem than they were for previous uses of these data
(Cioni \etal 2000b,c). We therefore reconsidered the issue of
zeropoint errors, and found that the inferred values $D_{ln}$ can be
used to successfully correct most of the zero-point errors in the
catalog.\footnote{Cioni \etal 2000a (their Section 3.2.3) already used
the $D_{ln}$, but only with the goal of correcting the few strips with
very poor calibrations, and not with the intention of improving the
calibration of the whole catalog. Note in this context that it was
reported there that one strip has a zeropoint error of $2.45$
magnitudes (which was manually corrected). This has now been traced to
a software bug that impacted only this particular strip; this bug has
now been fixed. The new calibration algorithm presented here shows
that necessary zeropoint corrections are always $\lta 0.25$ mag, with
$0.05$ mag being a more typical number.} To outline the approach, let
$z_l$ and $z_n$ be the true zeropoints that should (ideally) have been
used in the reduction of the strips with indices $l$ and $n$, and let
$Z_l$ and $Z_n$ be the zeropoints that were actually used in the data
reduction, based on the analysis of standard star observations. One
can define
\begin{equation}
\delta_l \equiv z_l - Z_l , \qquad 
\delta_n \equiv z_n - Z_n .
\label{deltadef}
\end{equation}
Non-zero values of $\delta_l$ and $\delta_n$ generally yield a shift
in the magnitude scale between the strips. The observationally
determined $D_{ln}$ provide a direct constraint on this shift:
\begin{equation}
-\delta_l + \delta_n = D_{ln} \pm \Delta D_{ln} .
\label{overlapdif}
\end{equation}
This constraint by itself is not sufficient to determine either
$\delta_l$ or $\delta_n$. As an example, if $D_{ln}$ is positive, one
does not know whether the zeropoint for strip $l$ should be decreased,
or the zeropoint for strip $n$ should be increased (or both). However,
equation~(\ref{overlapdif}) is not the only constraint that is
available on the values of $\delta_l$ and $\delta_n$. The
observationally determined zeropoints $Z_l$ and $Z_n$ have known
formal errors $\Delta Z_l$ and $\Delta Z_n$. Hence, one must have that
\begin{equation}
\delta_l = 0 \pm \Delta Z_l , \qquad
\delta_n = 0 \pm \Delta Z_n ,
\label{formalzperr}
\end{equation}
which uses the standard definition of error bars (i.e., $\delta_l$ and
$\delta_n$ are random deviates drawn from Gaussian distributions of
dispersions $\Delta Z_l$ and $\Delta Z_n$,
respectively). Equations~(\ref{overlapdif}) and~(\ref{formalzperr})
provide 3 linear equations for the two unknowns $\delta_l$ and
$\delta_n$. More generally, if there are $N$ adjacent strips, there
are $2N-1$ equations for the $N$ unknown values $\delta_1, \ldots,
\delta_N$. In the absence of other information, the optimum solution 
is the one that minimizes the chi-squared quantity
\begin{equation}
\chi^2 \equiv 
  \sum_{n=1}^{N-1} 
  \Bigl ( { {-\delta_n + \delta_{n+1} - D_{n,n+1}} \over
    {\Delta D_{n,n+1}} } \Bigr )^2 +
  \sum_{n=1}^{N}
  \Bigl ( { {\delta_n} \over {\Delta Z_n} } \Bigr )^2 .
\label{chisqcordef}
\end{equation}
The corresponding solution is easily obtained as the least-squares
solution of an overdetermined matrix equation. To obtain this solution
we used a singular value decomposition algorithm from Numerical
Recipes (Press \etal 1992). Having obtained the solution $\delta_1,
\ldots, \delta_N$, one can improve the zeropoint estimate $Z_n$ for each strip 
using equation~(\ref{deltadef}): $Z_n \rightarrow Z_n + \delta_n$, for
$n=1,\ldots,N$. This additional calibration was applied to the catalog
before performing the analysis of the present paper.  This improved
calibration will be made available through the CDS database in
Strasbourg together with the DCMC, so that it can be used in
subsequent analyses of this dataset.

The $\chi^2$ quantity in equation~(\ref{chisqcordef}) is a sum of two
terms. The first term measures how well the overlap regions between
different scan strips agree. This is the quantity that one would in
principle like to minimize. However, minimization of this term by
itself is an ill-conditioned problem. It corresponds to application of
a zeropoint offset to each scan strip to make it agree with the strip
next to it. It is easy to see that this will amplify noise in a
random-walk manner, and will introduce spurious large-scale
gradients. The role of the second term in the $\chi^2$ is to avoid
this. It forces the modifications $\delta$ to remain small, consistent
with the formal errors in the zeropoints, and hence acts as a
regularization term (e.g., Press \etal 1992). The minimum $\chi^2$
solution is therefore a compromise (in the usual sense of
regularization problems) that optimizes the agreement in the overlap
regions between different scan strips, but without noise amplification
and without the introduction of spurious large-scale gradients.

After applying the new calibration, one can again study the magnitude
differences for stars in the overlap region of adjacent strips. One
now obtains ${\rm RMS}(D_{ln})_I = 0.02$, ${\rm RMS}(D_{ln})_J =
0.02$, and ${\rm RMS}(D_{ln})_{K_s} = 0.03$. As discussed above, the
zeropoint errors per strip can be estimated as $1/\sqrt{2}$ times
these numbers; this yields error estimates of $0.01$--$0.02$
magnitudes.  Having said this, it should be kept in mind that at the
level of hundredths of a magnitude there may be effects other than
zeropoint errors between strips that could be impacting the accuracy
of the catalog. In particular, there could be zeropoint variations of
this order {\it within} a given strip (each strip, after all, is made
up of 180 separate but overlapping images). The only information that
is available on such variations comes from the variation of the
magnitude differences for stars in overlap regions as function of
declination. For the large majority of the strips such variations are
minimal, and do not exceed the level of a few hundredth of a
magnitude. However, there are a few strips where larger variations are
present, presumably as a result of sub-optimal observing
conditions. It was verified that none of these strips (less than a
handful) had a significant impact on the results presented in this
paper.

In summary, the DCMC, especially with the improved calibration, allows
highly accurate studies of the positional-dependence of features in
the near-IR CMDs. The results of the present study provide direct
post-hoc confirmation of this, since the RMS residuals in data-model
comparisons such as that of Figure~\ref{f:fits} are no larger than
$\sim 0.03$ magnitudes. Also, the DENIS results agree at this same
level of accuracy with the independent analysis of 2MASS data
(cf.~Figure~\ref{f:banddep}).

It should be kept in mind that none of the above discussion refers to
a possible overall {\it absolute} zeropoint error in the whole
catalog. Such errors could, e.g., result from possible errors in the
absolute calibration of the DENIS photometric passbands (Fouqu\'e
\etal 2000), although there is currently no reason to believe that there
are significant errors of this kind. If present, absolute errors would
certainly impact distance determinations of the LMC. However, they
would not affect the analysis presented here, which only uses {\it
relative} magnitudes and distances of different areas of the LMC.


\ifsubmode\else
\baselineskip=10pt
\fi


\clearpage


\ifsubmode\else
\baselineskip=14pt
\fi


\newcommand{\figcapmaxminvar}
{Solid curves show the magnitude variation $\mu$ (defined by
equation~(\ref{magdif})) for points that are observed perpendicular to
the line of nodes (i.e., $\phi = \theta \pm 90^{\circ}$, where
$\theta$ is the position angle of the line of nodes) for a plane
inclined at angle $i$ with respect to the plane of the sky. The
coordinates $(\rho,\phi)$ are angular coordinates on the
sky. Predictions are shown for different values of the inclination
angle, as indicated in the figure. For comparison, the heavy
long-dashed line shows the linear Taylor approximation given by
equation~(\ref{dmagtaylor}), for the case $i=40^{\circ}$. The label
along the abscissa is $\rho \sin (\phi - \theta)$, where by assumption
$\sin (\phi - \theta) = \pm 1$. Points with $\rho \sin (\phi - \theta)
> 0$ are tilted away from the observer, and points with $\rho \sin
(\phi - \theta) < 0$ are tilted towards the
observer.\label{f:maxminvar}}

\newcommand{\figcapCMDs}
{The panels show the nine independent CMDs that can be constructed
from the $I$, $J$ and $K_S$ magnitudes obtained from the DENIS Survey.
Each panel shows all the stars in the LMC area of the sky covered by
the DCMC that were detected in all 3-bands. The features in the CMDs
are discussed in the text. The horizontal bar at the right axis of
each panel indicates the magnitude of the Tip of the Red Giant Branch
(TRGB), as determined by Cioni \etal (2000c). The vertical lines in
the top left panel indicate the range of $I-J$ colors that will be
used for the main analysis of the present paper. The $I$-band LF of
the stars with magnitudes in this color range is shown in
Figure~\ref{f:histograms}. The vertical lines in the bottom left panel
show a range of $J-K_s$ colors that is used as a consistency check in
Section~\ref{s:difbands}.\label{f:CMDs}}

\newcommand{\figcaphistograms}
{$I$-band Luminosity Function $N(I)$ extracted from the DCMC for stars
with $1.6 \leq I-J \leq 2.0$ (as indicated in the top-left panel of
Figure~\ref{f:CMDs}). The ordinate is the number of stars per $0.15$
mag wide bin. The peak is due to AGB stars. The tail towards faint
magnitudes is due RGB stars that are intrinsically bluer than $I-J =
1.6$, but for which the observed color is in the range $1.6 \leq I-J
\leq 2.0$ due to photometric errors. These stars do not affect the peak 
in the LF, because the tip of the RGB is at $I=14.5$ (Cioni \etal
2000c).\label{f:histograms}}

\newcommand{\figcapimagegrid}
{Schematic representation of the LMC area of the sky. The $(X,Y)$
coordinates are the coordinates of a `zenithal equidistant
projection', as defined by equation~(\ref{projdef}). Dotted curves are
curves of constant RA and declination, as labeled at the top and right
of the figure, respectively. The area outlined by the long-dashed
heavy curves is the region covered by the DCMC. Solid heavy curves
show the polar grid that subdivides the outer area of the LMC in
different sectors used for the analysis presented
here.\label{f:imagegrid}}
 
\newcommand{\figcapfits}
{Results of the analysis of the $I$-band LF of stars selected from the
DCMC to have $1.6 \leq I-J \leq 2.0$. The quantity $\mu$ measures the
variation in the AGB modal magnitude along a ring on the sky. Each
panel corresponds to a different ring, as indicated, with the
innermost ring in the top panel and the outermost ring in the bottom
panel. The position angle $\Phi$ is plotted along the abscissa, with
each datapoint corresponding to a different azimuthal sector
(cf.~Figure~\ref{f:imagegrid}). The curves show the best model fits to
the data. The solid curves show the results when the viewing angles
are allowed to be different for each radial ring, while the dashed
curves show the results when only a single combination of viewing
angles is allowed for all rings (i.e., no warps or twists in the LMC
plane). The values of $(i,\Theta)$ for these models are shown in the
top panels of Figure~\ref{f:viewang}.\label{f:fits}}

\newcommand{\figcapbanddep}
{Results of several independent analyses for the single ring
$2.5^{\circ} \leq \rho \leq 6.7^{\circ}$, subdivided in $N=8$
azimuthal sectors. The position angle $\Phi$ is plotted along the
abscissa, and the variation $\mu$ in apparent magnitude along the ring
is plotted along the ordinate. {\bf (Top panel)} Modal magnitude of
(AGB) stars in the DCMC with $1.6 \leq I-J \leq 2.0$.  {\bf (Second
panel)} Modal magnitude of stars in the DCMC with $1.5 \leq J-K_s \leq
2.0$. {\bf (Third panel)} TRGB magnitude of stars in the DCMC. {\bf
(Bottom panel)} Modal magnitude of stars in the 2MASS Point Source
Catalog with $1.5 \leq J-K_s \leq 2.0$. Filled circles are results
from the $I$-band LF (DENIS data only), open circles are results from
the $J$-band LF, four-pointed stars are results from the $H$-band LF
(2MASS data only), and open triangles are results from the $K_s$-band
LF. Results in different bands are plotted with small horizontal
offsets to avoid confusion. The results from the different methods, in
the different photometric bands, and from the different surveys are
all fully consistent. The dashed curve (identical in each panel) shows
the predictions for the model with $i = 34.7^{\circ}$ and $\Theta =
122.5^{\circ}$.\label{f:banddep}}

\newcommand{\figcapmzero}
{Variation $m_0 - \langle m_0 \rangle$ in the distance-corrected AGB
modal magnitude as function of angular distance $\rho$ from the LMC
center. The three panels show the results derived from $I$, $J$ and
$K_s$ band LFs, respectively. Filled points are for the color
selection criterion $1.6 \leq I-J \leq 2.0$, and open points for the
criterion $1.5 \leq J-K_s \leq 2.0$. The quantity $m_0$ is the AGB
modal magnitude for a single ring on the sky, while $\langle m_0
\rangle$ is the average of the $m_0$ values for all rings. Radial
variations in $m_0 - \langle m_0 \rangle$ could be due to variations
in, e.g., dust absorption or the age or the metallicity of the stellar
population. The observed variations are much smaller than the
azimuthal variations in the AGB modal magnitude (e.g.,
Figure~\ref{f:fits}) which can be attributed to distance
effects.\label{f:mzero}}

\newcommand{\figcapviewang}
{Dependence of the inclination angle $i$ and the position angle of the
line of nodes $\Theta$ on the angular distance $\rho$ from the LMC
center. Filled points show the results of the fit when the viewing
angles are allowed to be different for each radial ring (with
horizontal error bars indicating the size of each ring). Open points
with dotted error bars show the results when only a single combination
of viewing angles is allowed for all rings (with the horizontal error
bars indicating the full radial range of the study). {\bf (Top
panels)} Fits to results obtained from the $I$-band LF of AGB stars
selected by $I-J$ color from the DENIS survey. {\bf (Middle panels)}
Fits to results obtained from the the $J$, $H$ and $K_s$ band LFs of
AGB stars selected by $J-K_s$ color from the 2MASS survey. {\bf
(Bottom panels)} Fits to the combined results from all analyses that
were performed, as described in the text. The results with the highest
accuracy (bottom panels) suggest that both $i$ and $\Theta$ decrease
as function of $\rho$, possibly indicative of a warp in the LMC disk
plane.\label{f:viewang}}


\ifsubmode
\figcaption{\figcapmaxminvar}
\figcaption{\figcapCMDs}
\figcaption{\figcaphistograms}
\figcaption{\figcapimagegrid}
\figcaption{\figcapfits}
\figcaption{\figcapbanddep}
\figcaption{\figcapmzero}
\figcaption{\figcapviewang}

\clearpage
\else\printfigtrue\fi

\ifprintfig


\clearpage
\begin{figure}
\centerline{\epsfbox{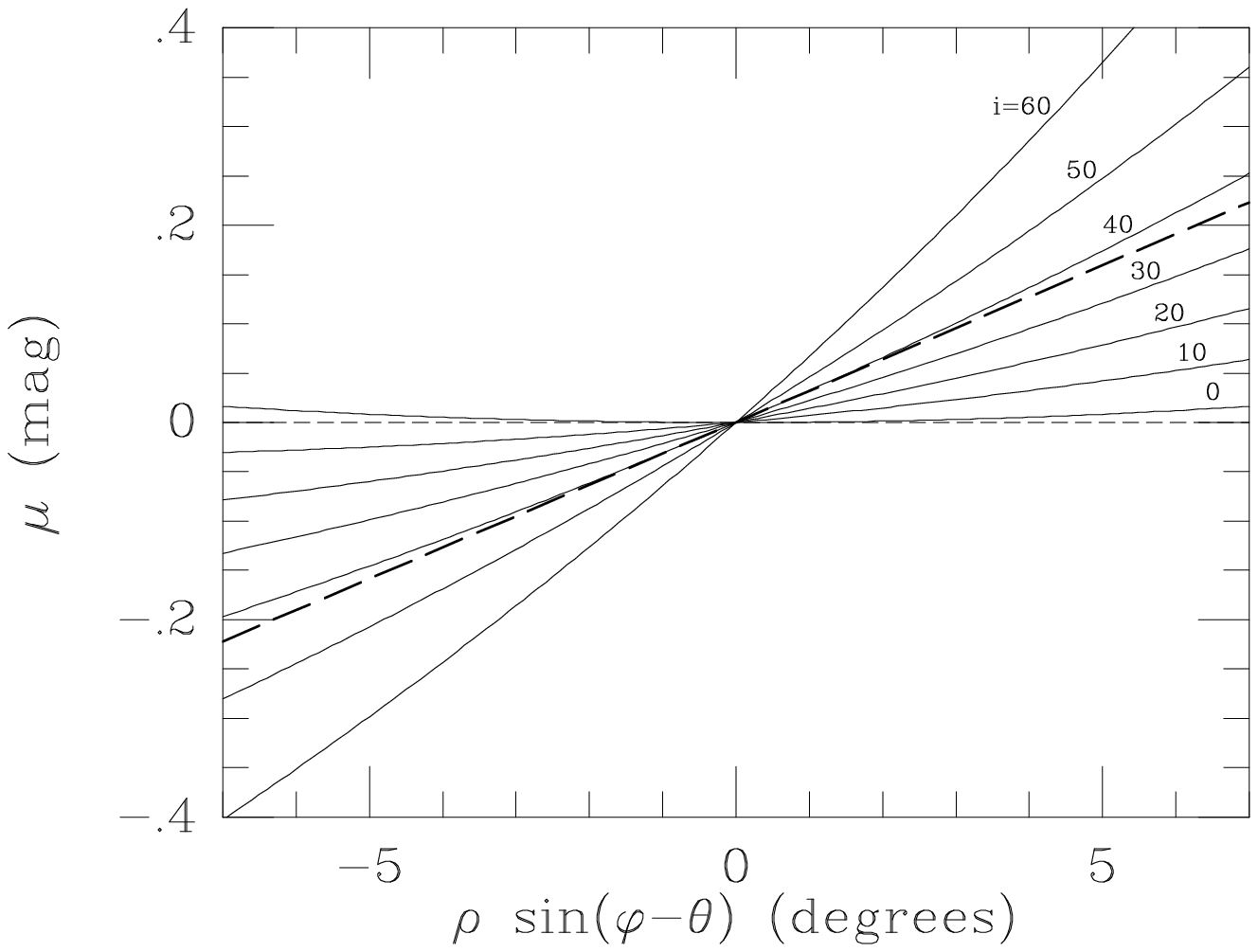}}
\ifsubmode
\vskip3.0truecm
\setcounter{figure}{0}
\addtocounter{figure}{1}
\centerline{Figure~\thefigure}
\else\figcaption{\figcapmaxminvar}\fi
\end{figure}


\clearpage
\begin{figure}
\centerline{%
\epsfxsize=0.3\hsize
\epsfbox{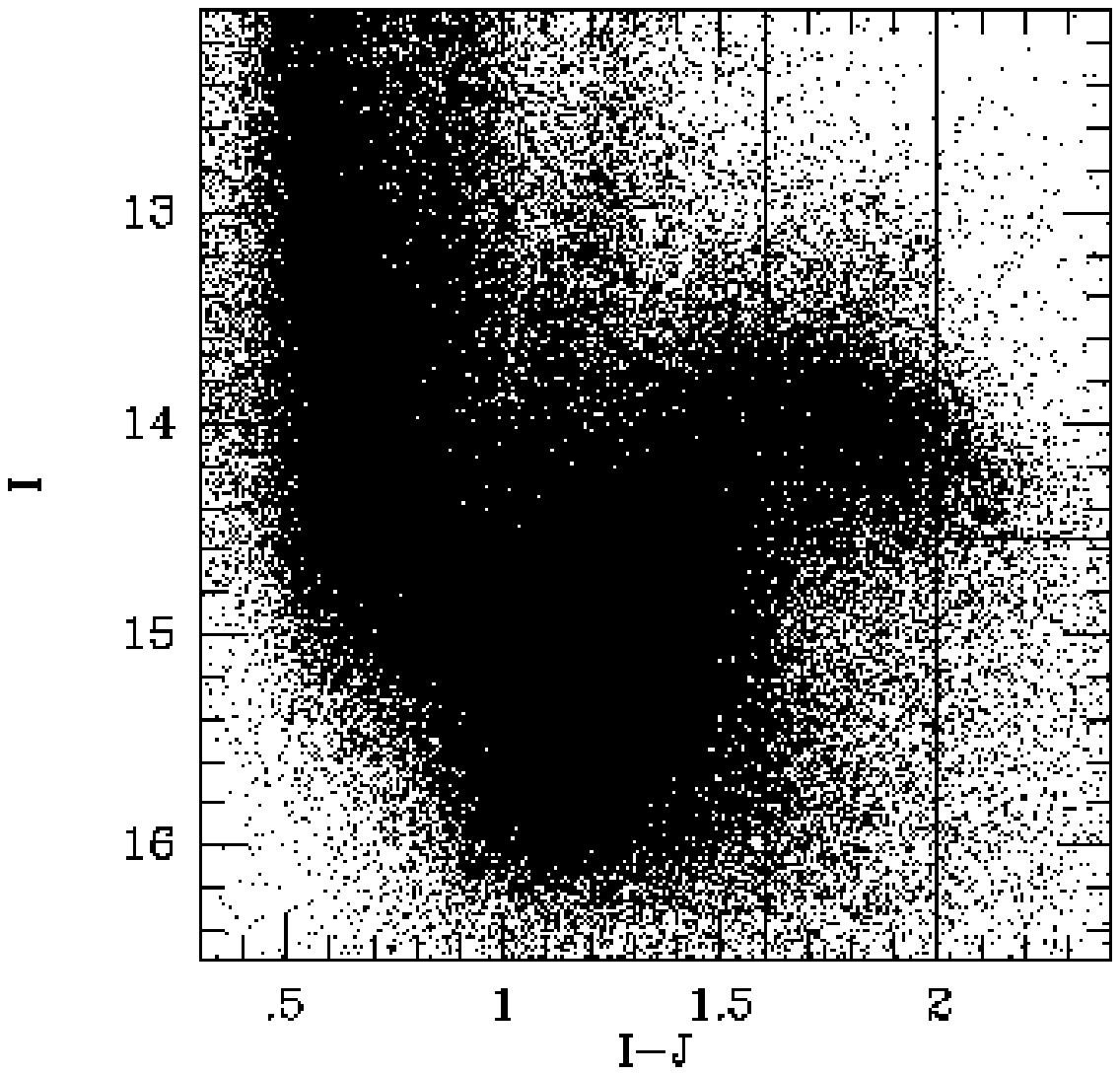}\quad
\epsfxsize=0.3\hsize
\epsfbox{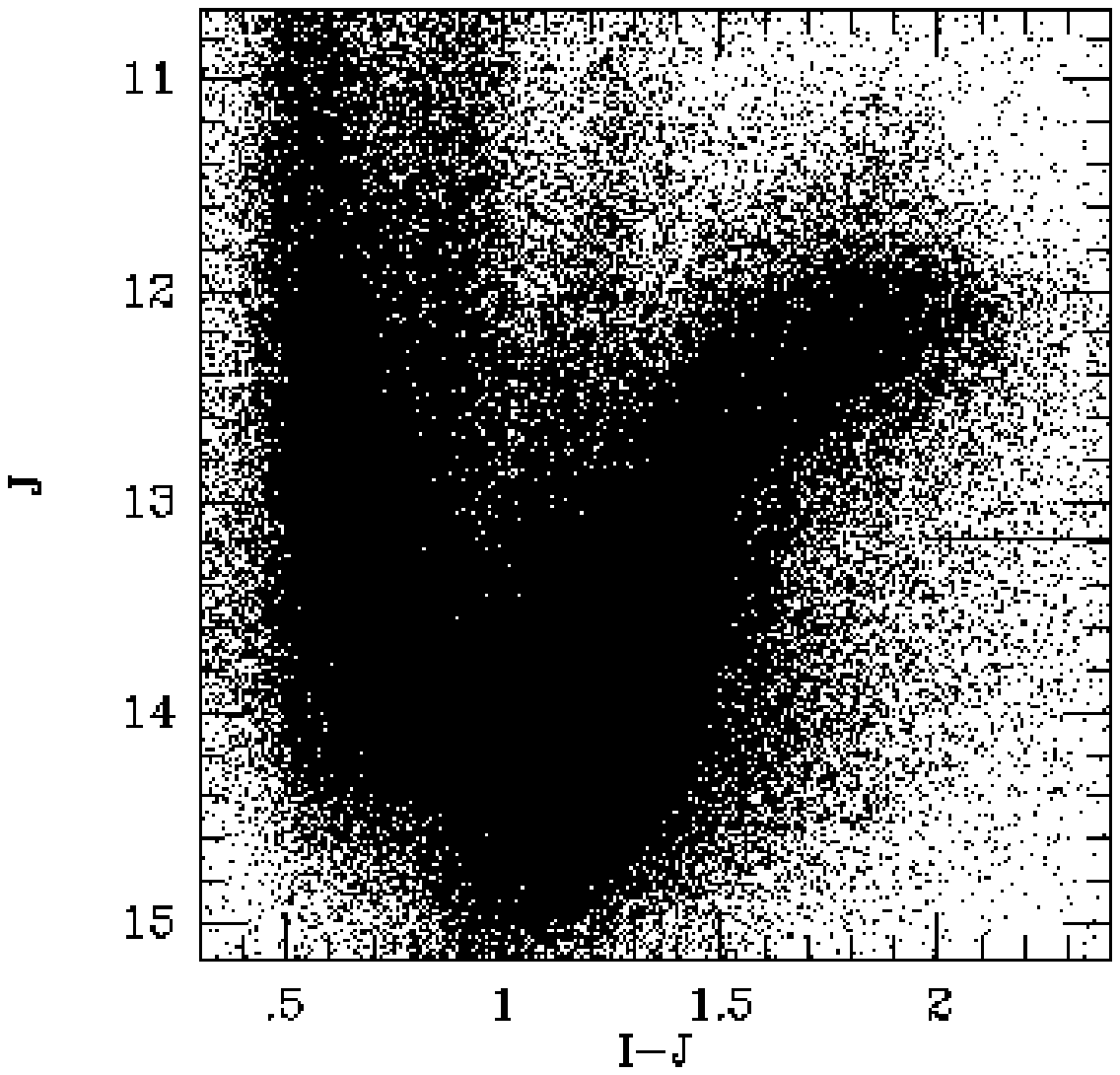}\quad
\epsfxsize=0.3\hsize
\epsfbox{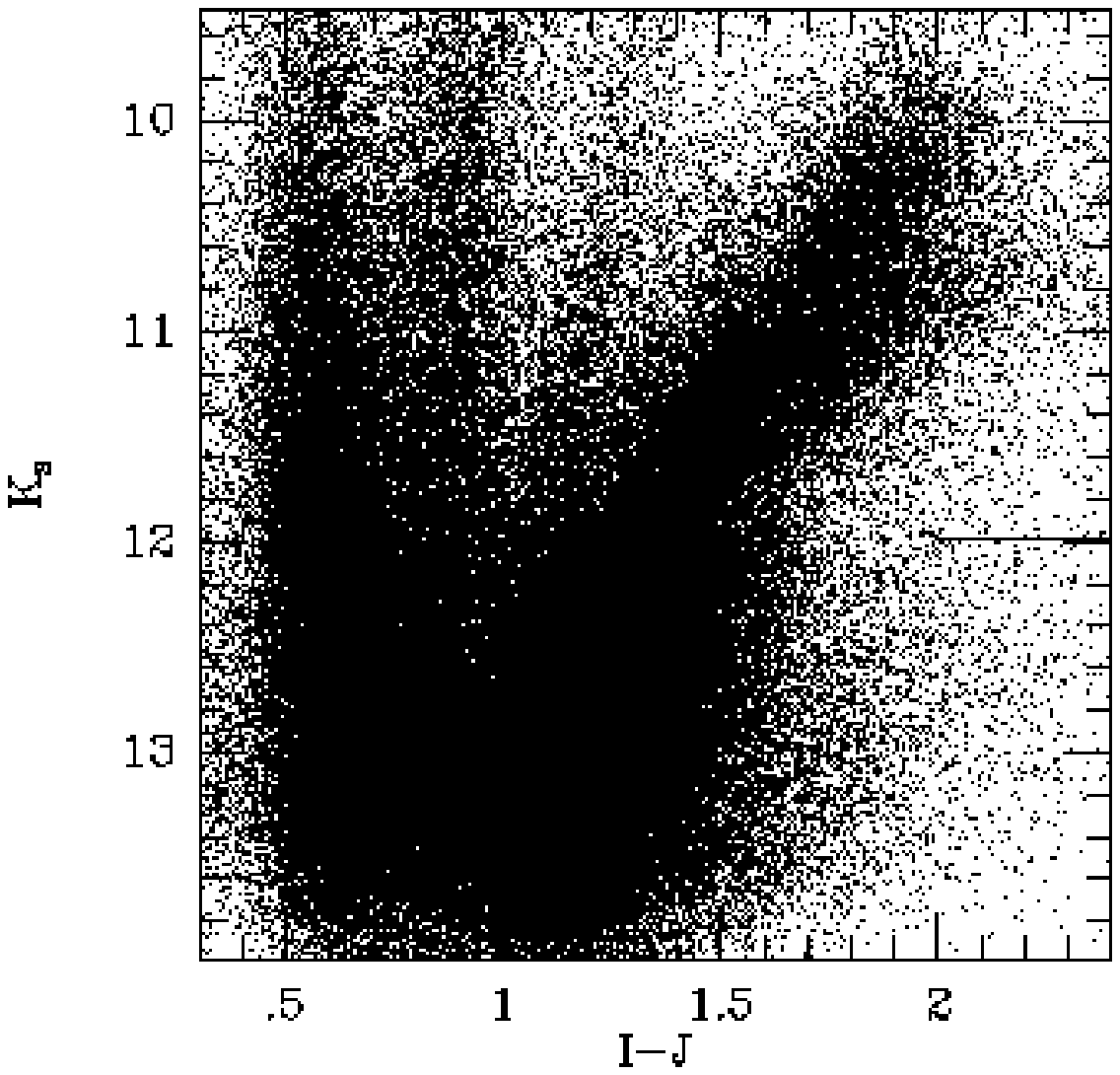}}
\smallskip
\centerline{%
\epsfxsize=0.3\hsize
\epsfbox{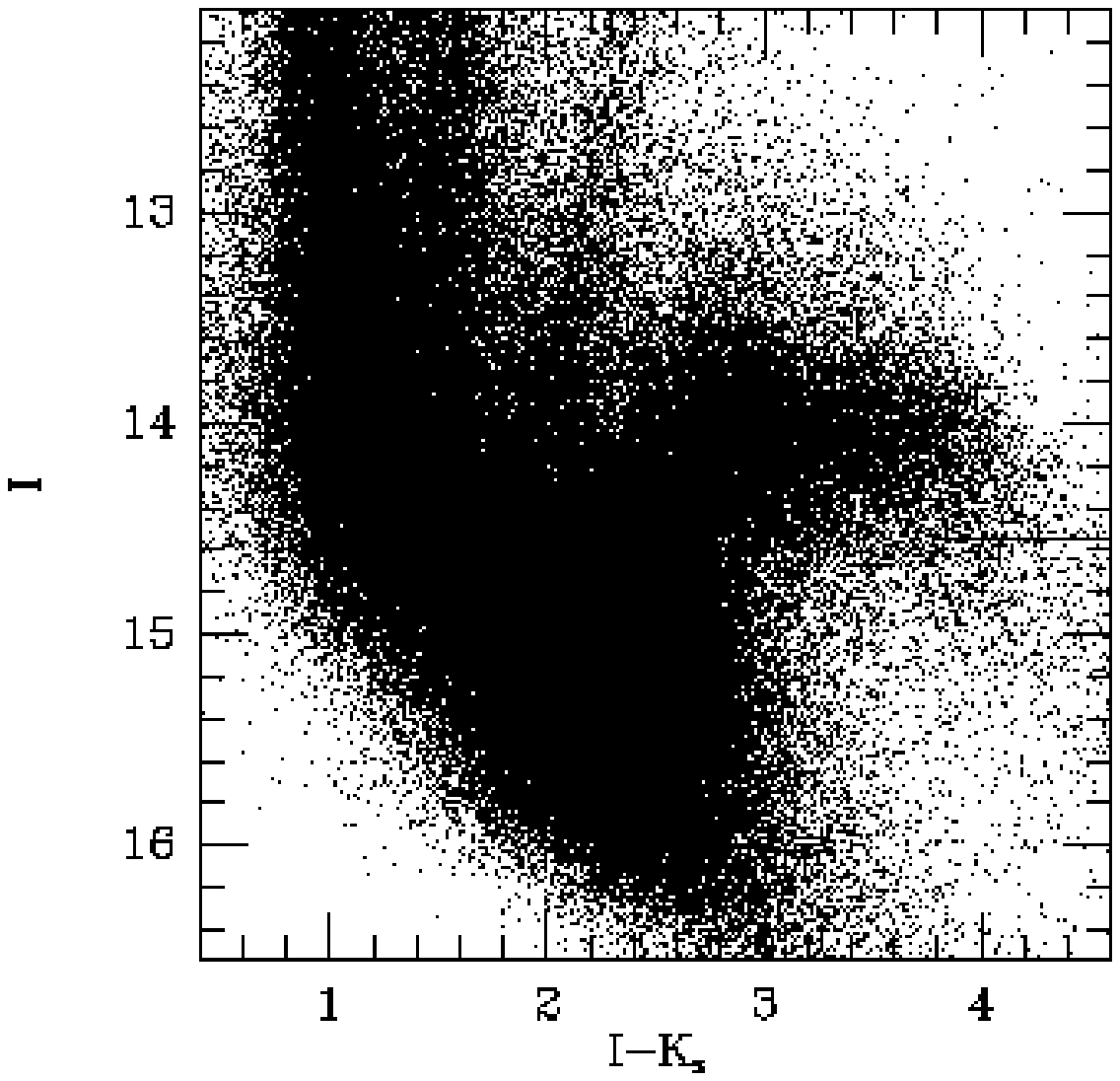}\quad
\epsfxsize=0.3\hsize
\epsfbox{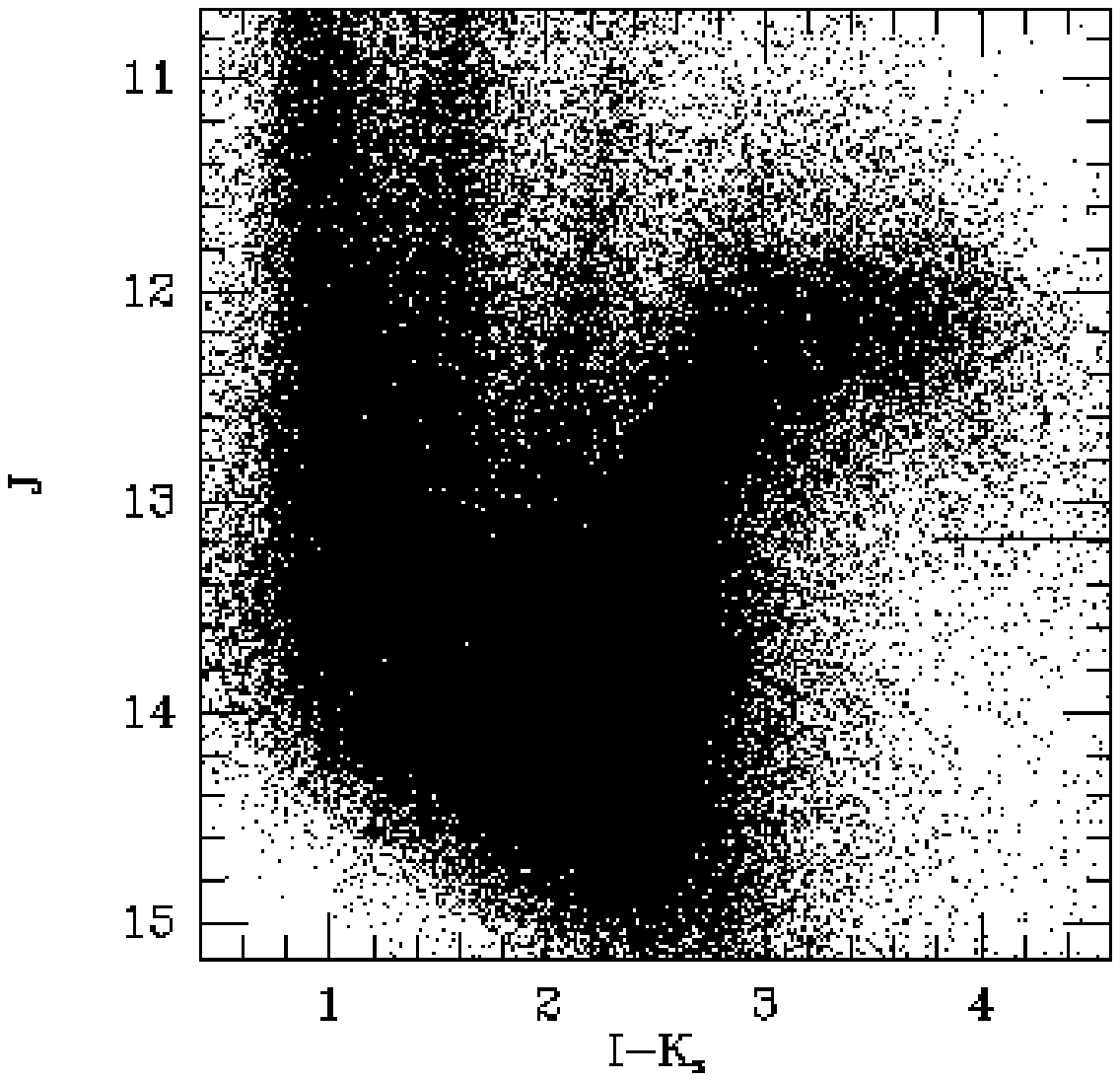}\quad
\epsfxsize=0.3\hsize
\epsfbox{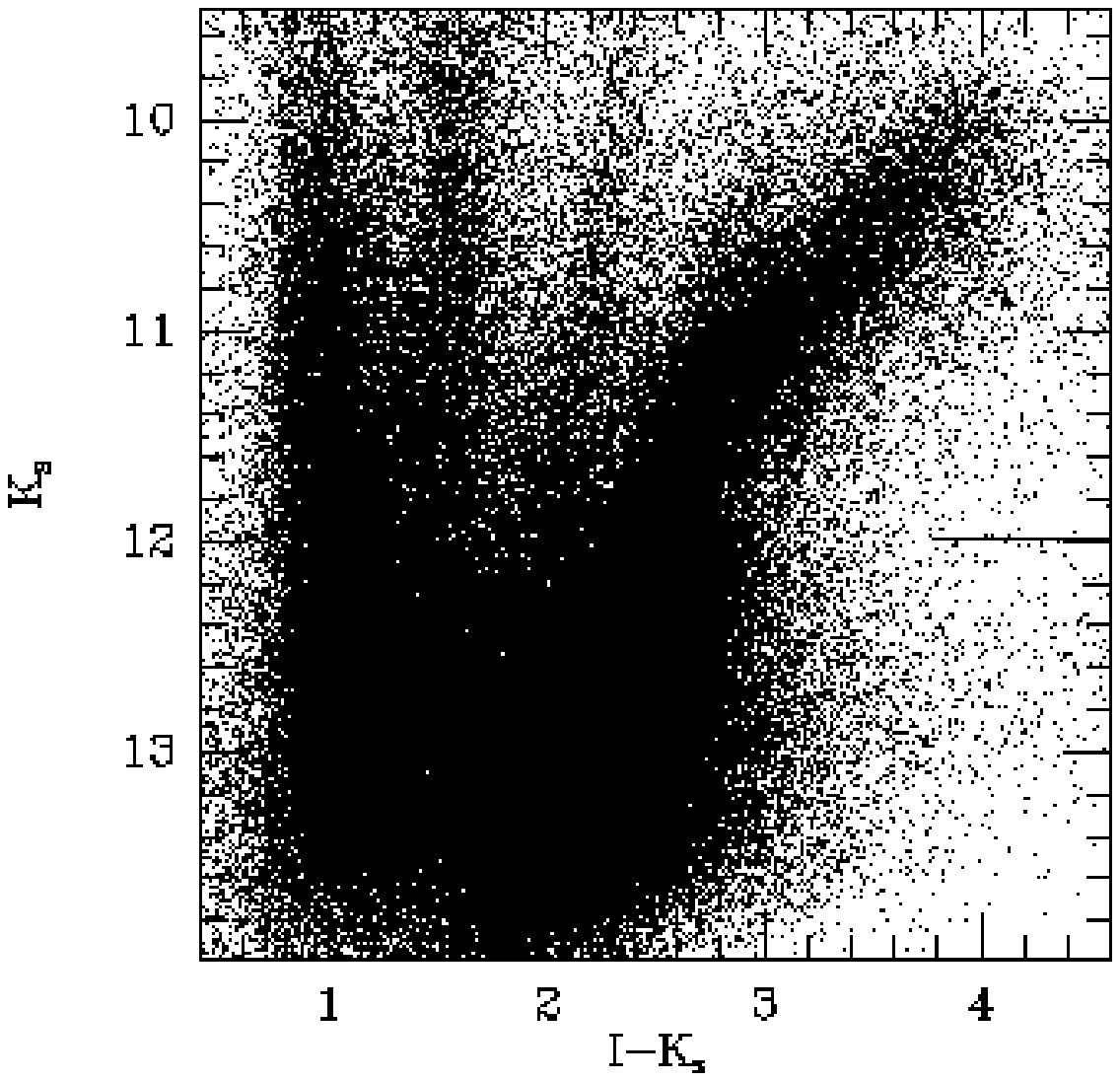}}
\smallskip
\centerline{%
\epsfxsize=0.3\hsize
\epsfbox{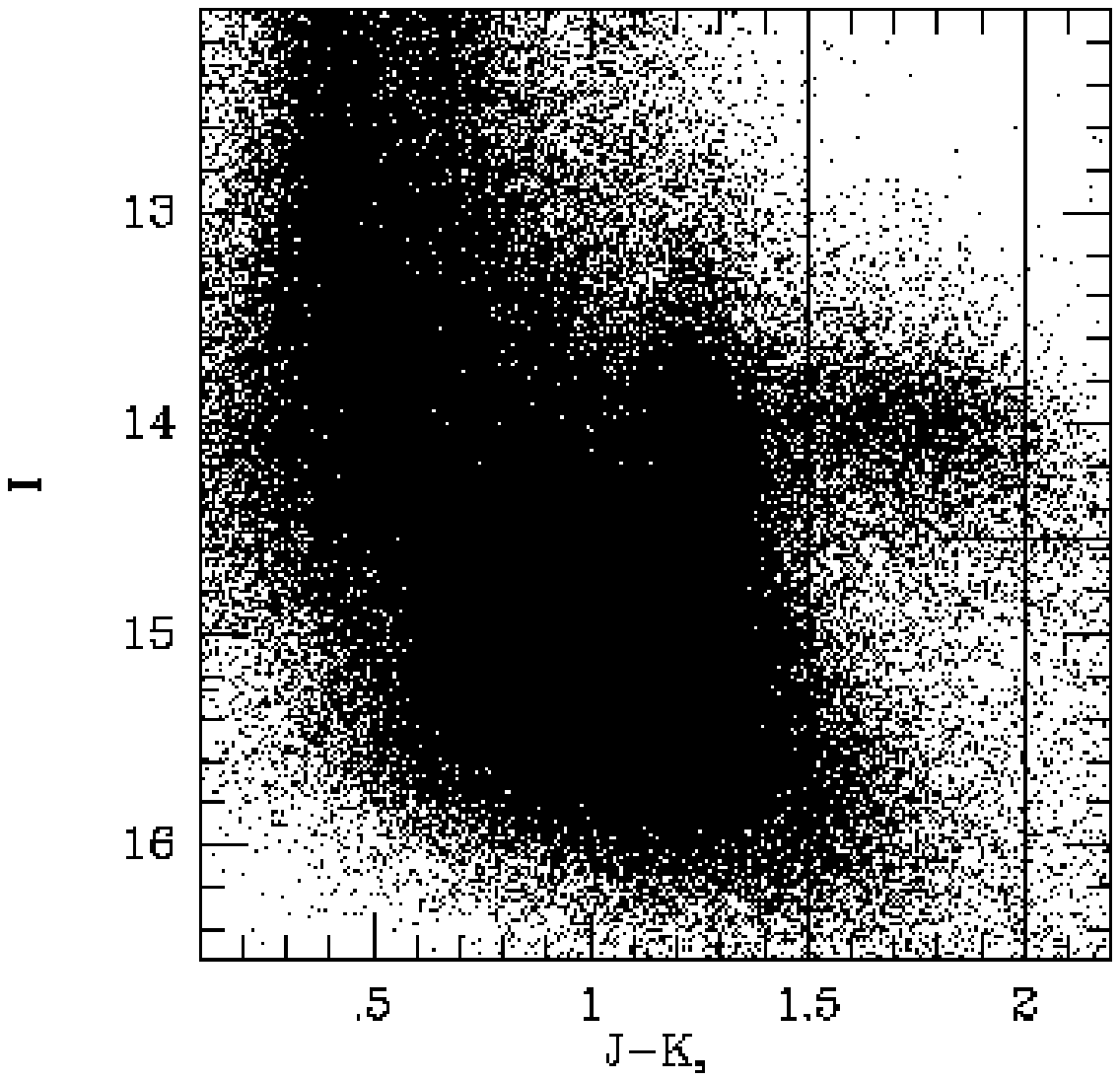}\quad
\epsfxsize=0.3\hsize
\epsfbox{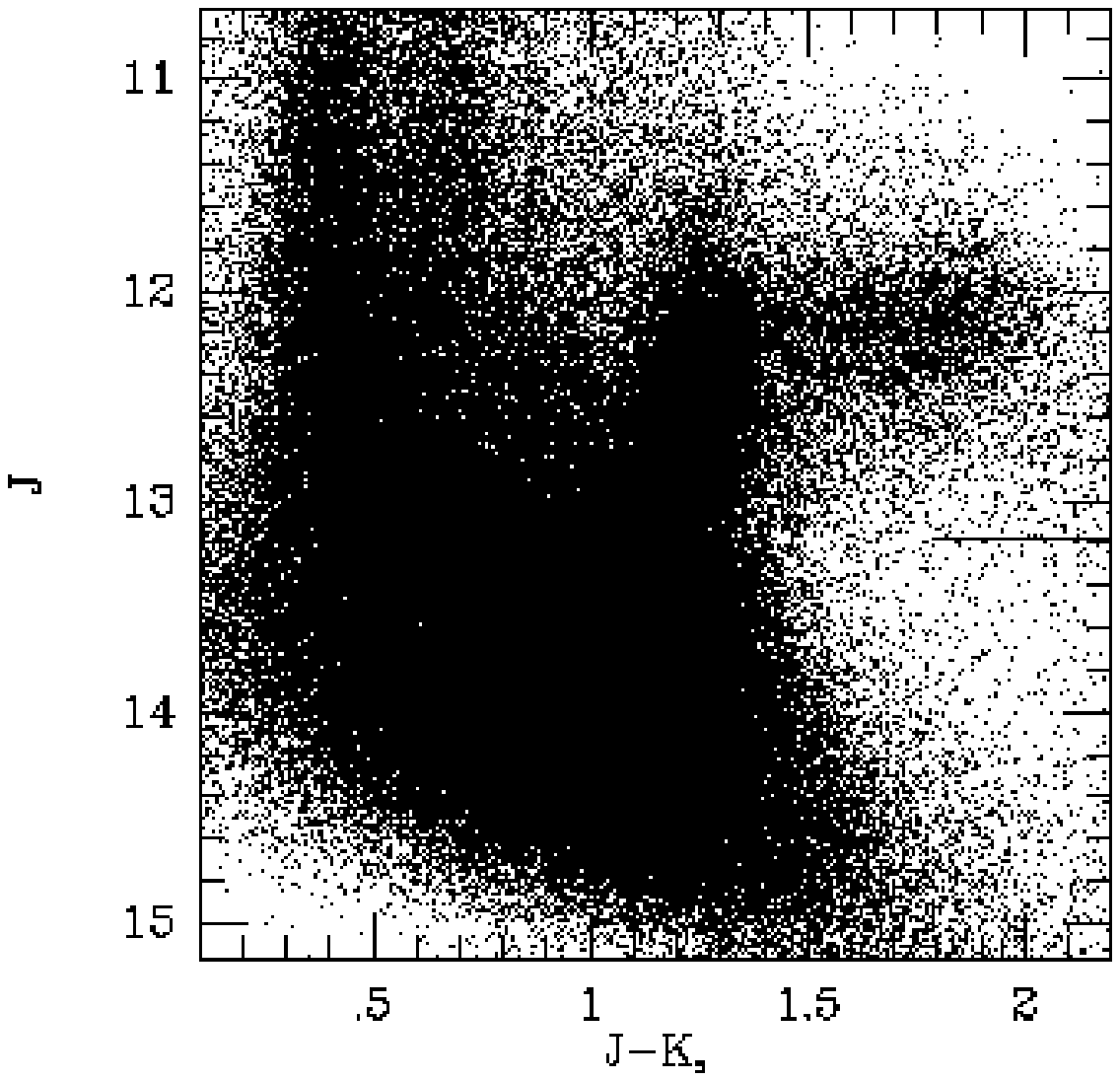}\quad
\epsfxsize=0.3\hsize
\epsfbox{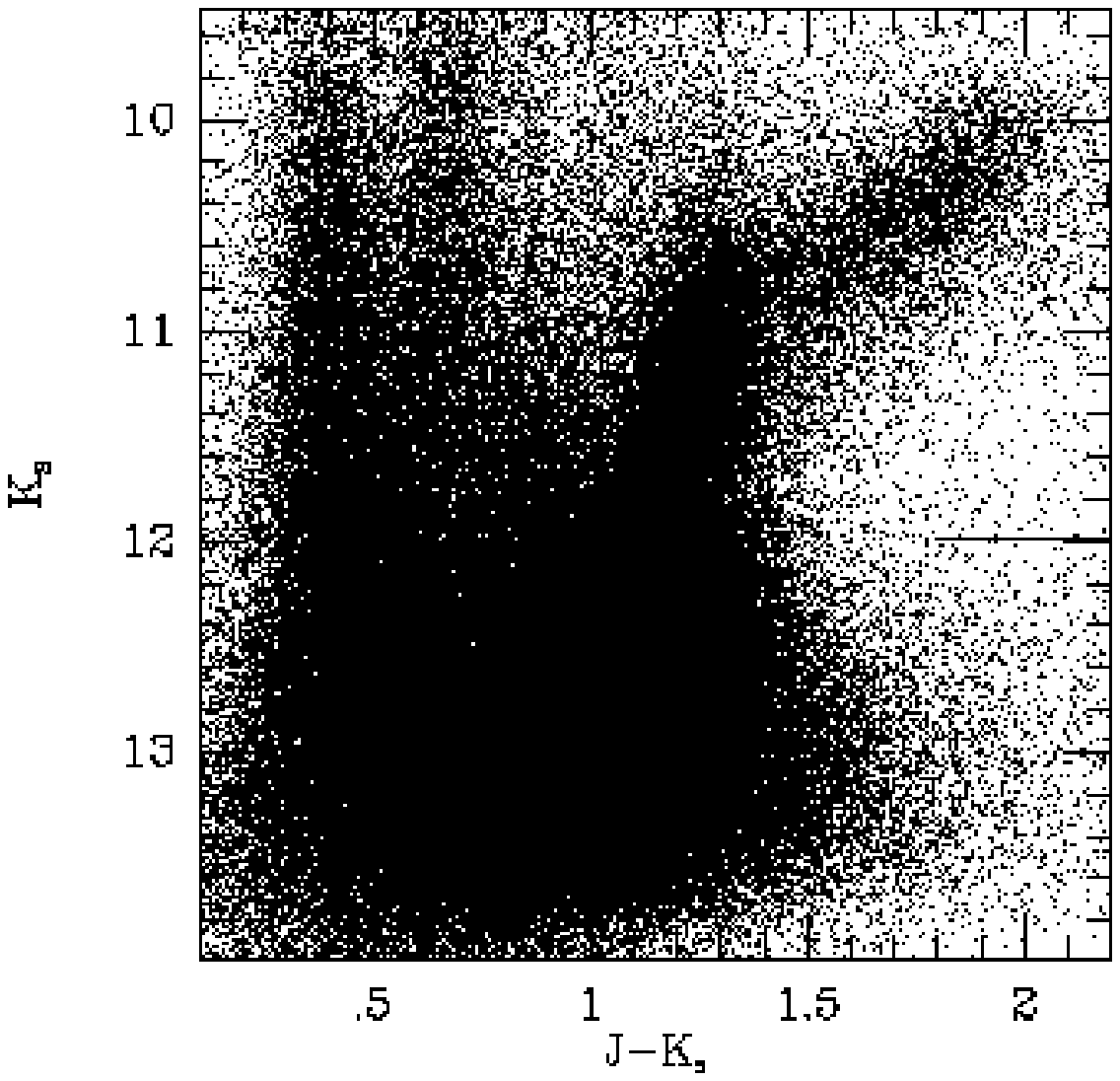}}
\ifsubmode
\vskip3.0truecm
\addtocounter{figure}{1}
\centerline{Figure~\thefigure}
\else\figcaption{\figcapCMDs}\fi
\end{figure}


\clearpage
\begin{figure}
\epsfxsize=0.7\hsize
\centerline{\epsfbox{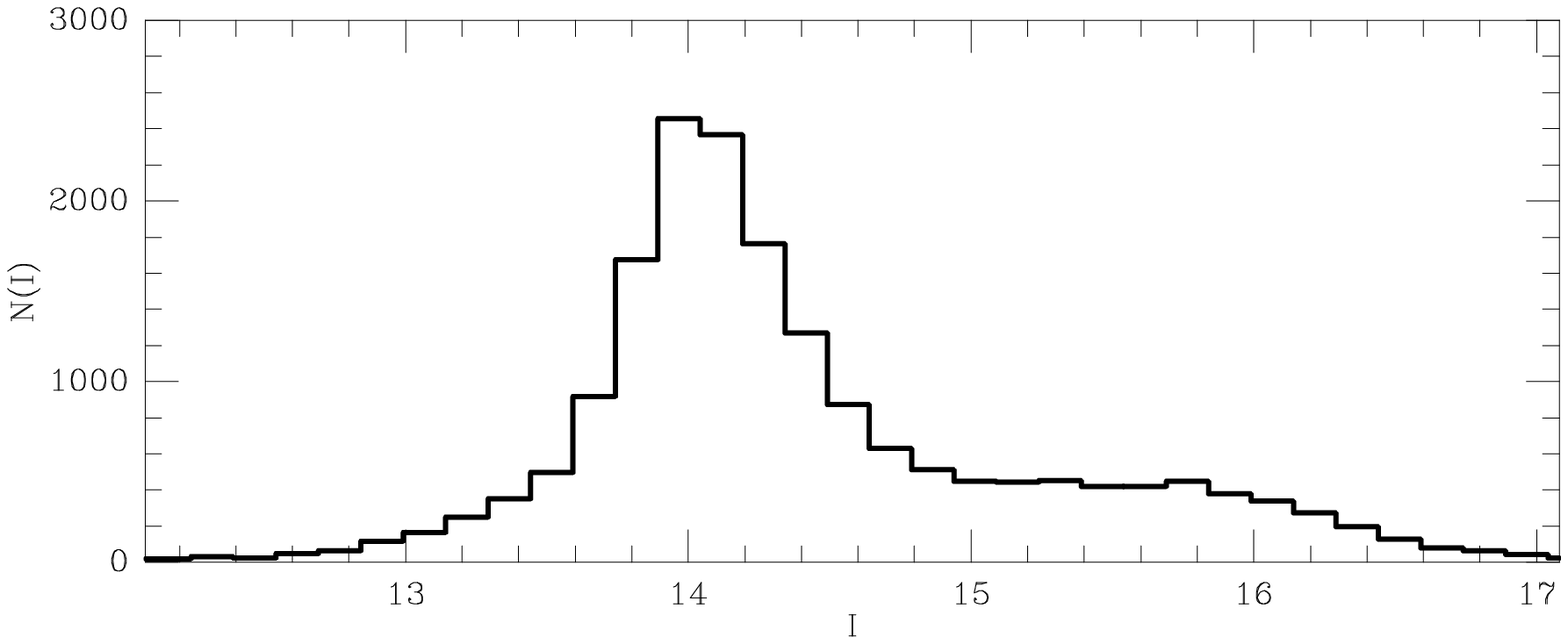}}
\ifsubmode
\vskip3.0truecm
\addtocounter{figure}{1}
\centerline{Figure~\thefigure}
\else\figcaption{\figcaphistograms}\fi
\end{figure}


\clearpage
\begin{figure}
\epsfxsize=0.9\hsize
\centerline{\epsfbox{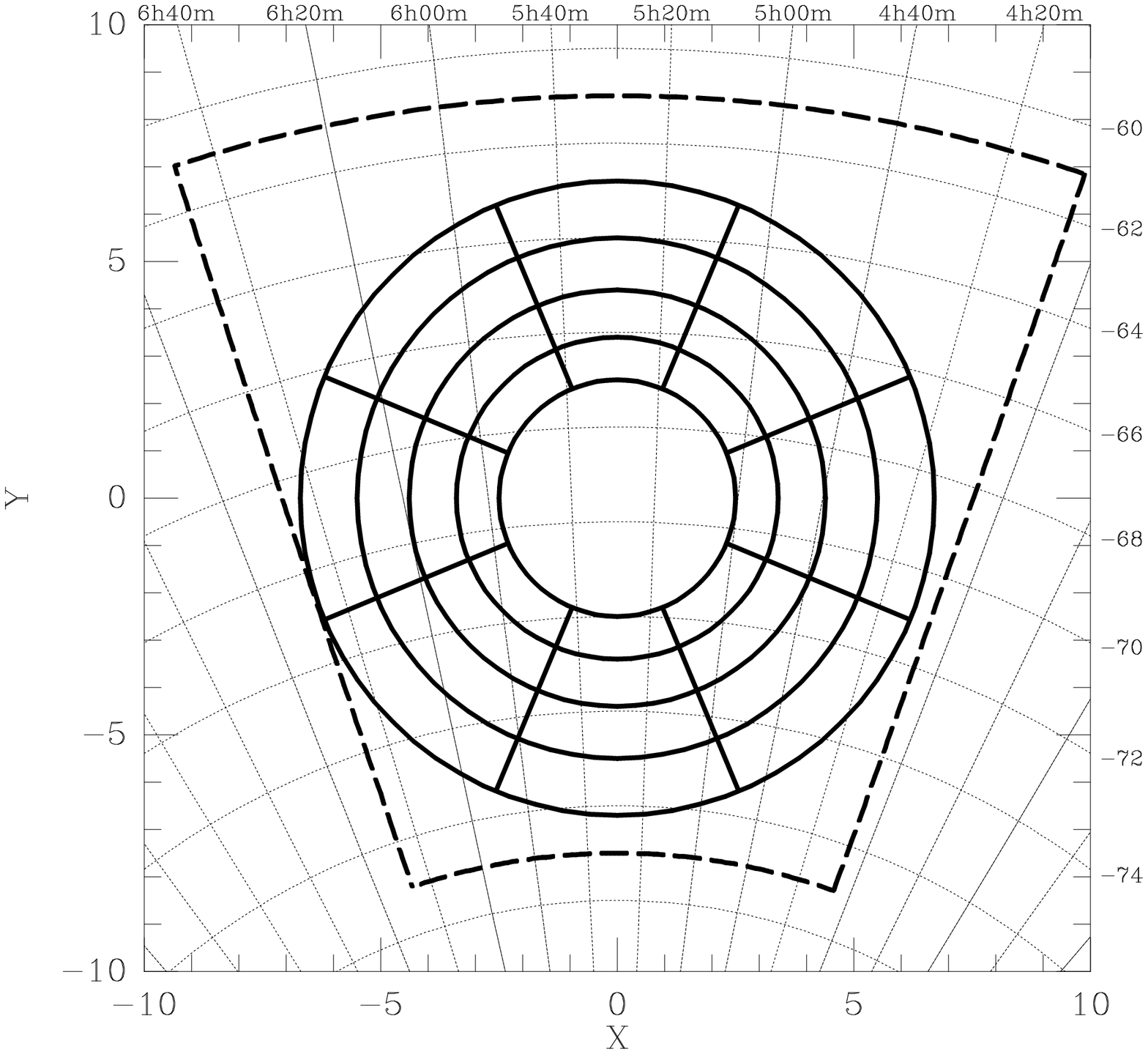}}
\ifsubmode
\vskip3.0truecm
\addtocounter{figure}{1}
\centerline{Figure~\thefigure}
\else\figcaption{\figcapimagegrid}\fi
\end{figure}


\clearpage
\begin{figure}
\epsfxsize=0.5\hsize
\centerline{\epsfbox{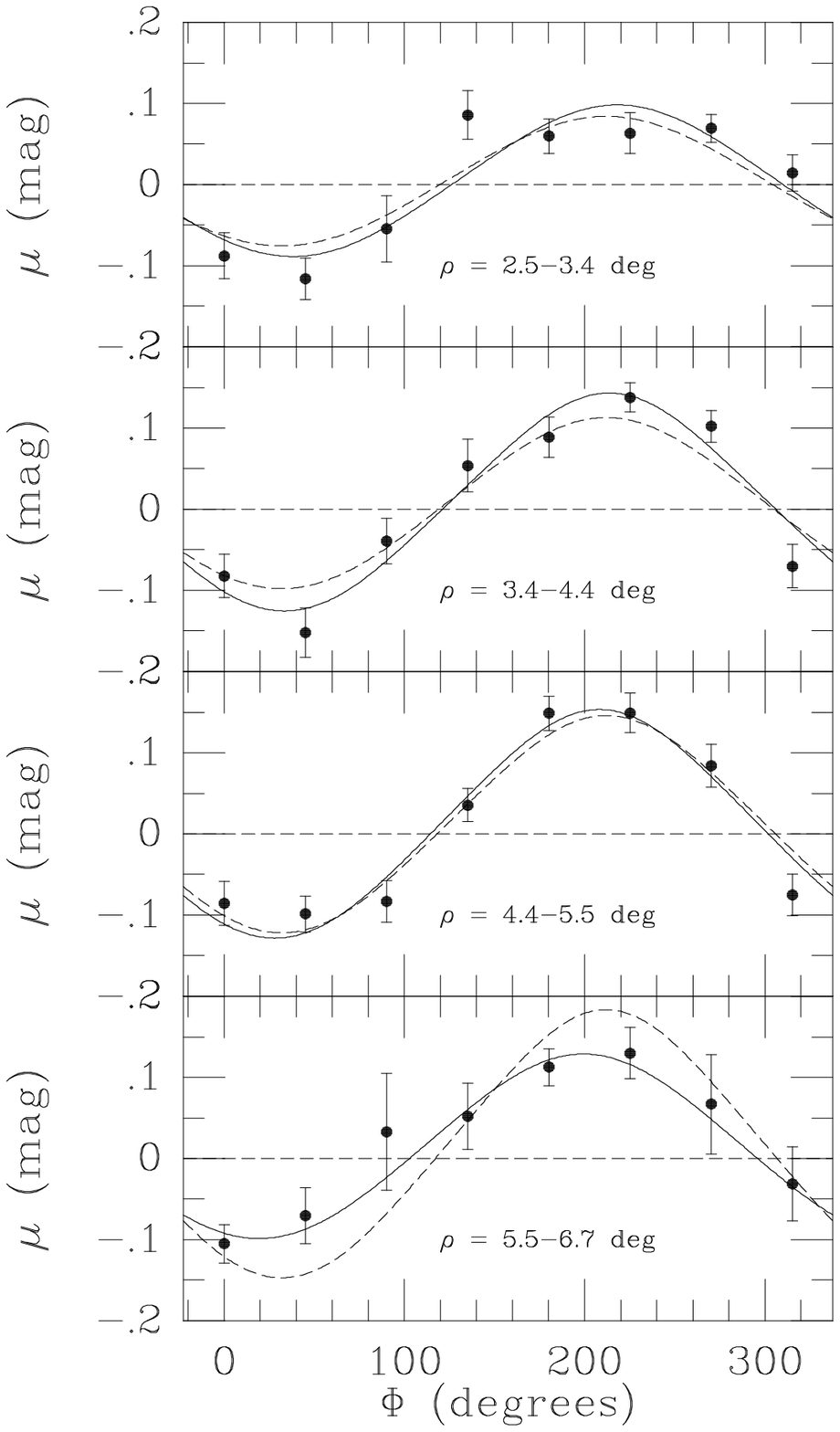}}
\ifsubmode
\vskip3.0truecm
\addtocounter{figure}{1}
\centerline{Figure~\thefigure}
\else\figcaption{\figcapfits}\fi
\end{figure}


\clearpage
\begin{figure}
\epsfxsize=0.5\hsize
\centerline{\epsfbox{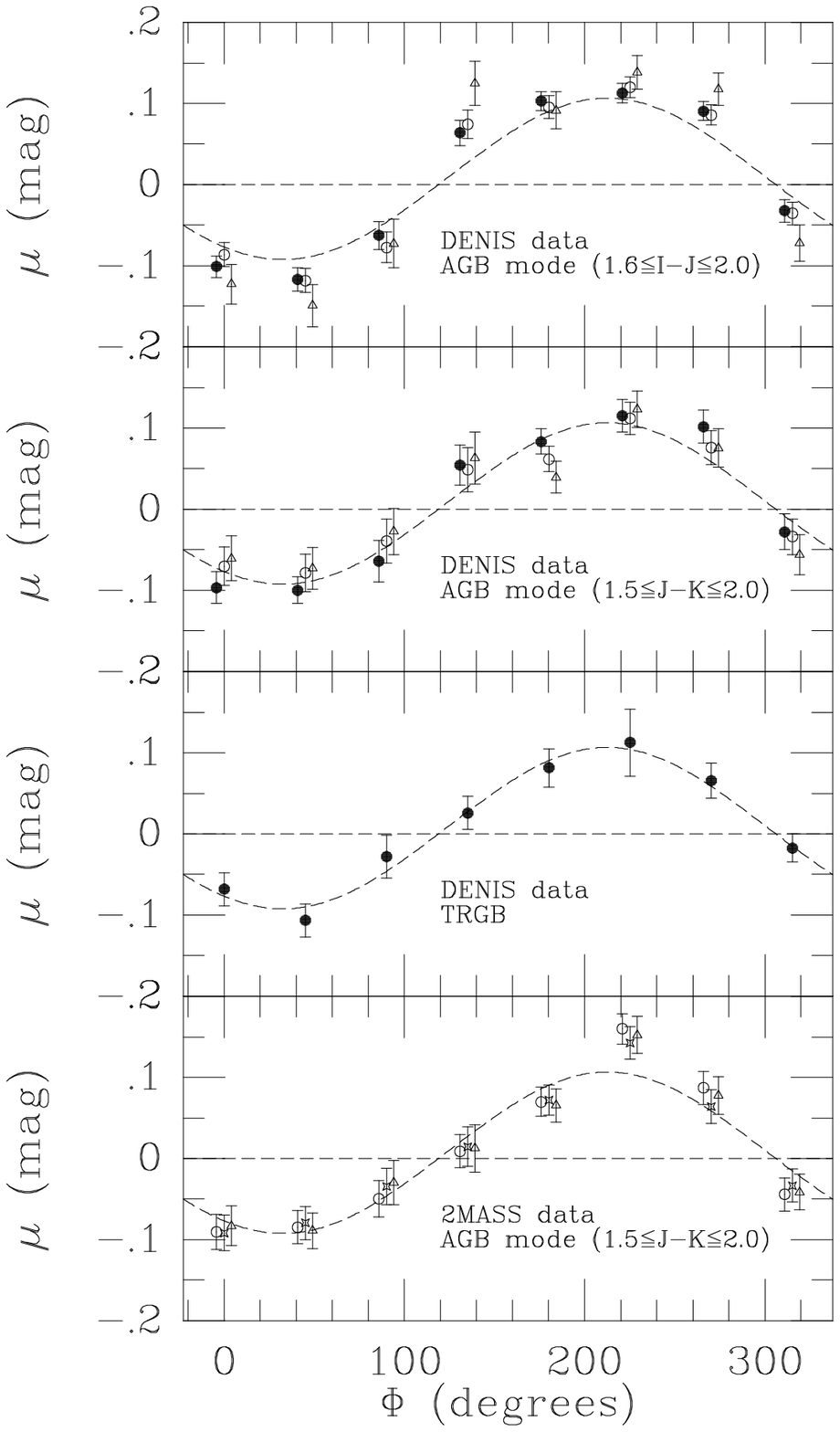}}
\ifsubmode
\vskip3.0truecm
\addtocounter{figure}{1}
\centerline{Figure~\thefigure}
\else\figcaption{\figcapbanddep}\fi
\end{figure}


\clearpage
\begin{figure}
\epsfxsize=0.9\hsize
\centerline{\epsfbox{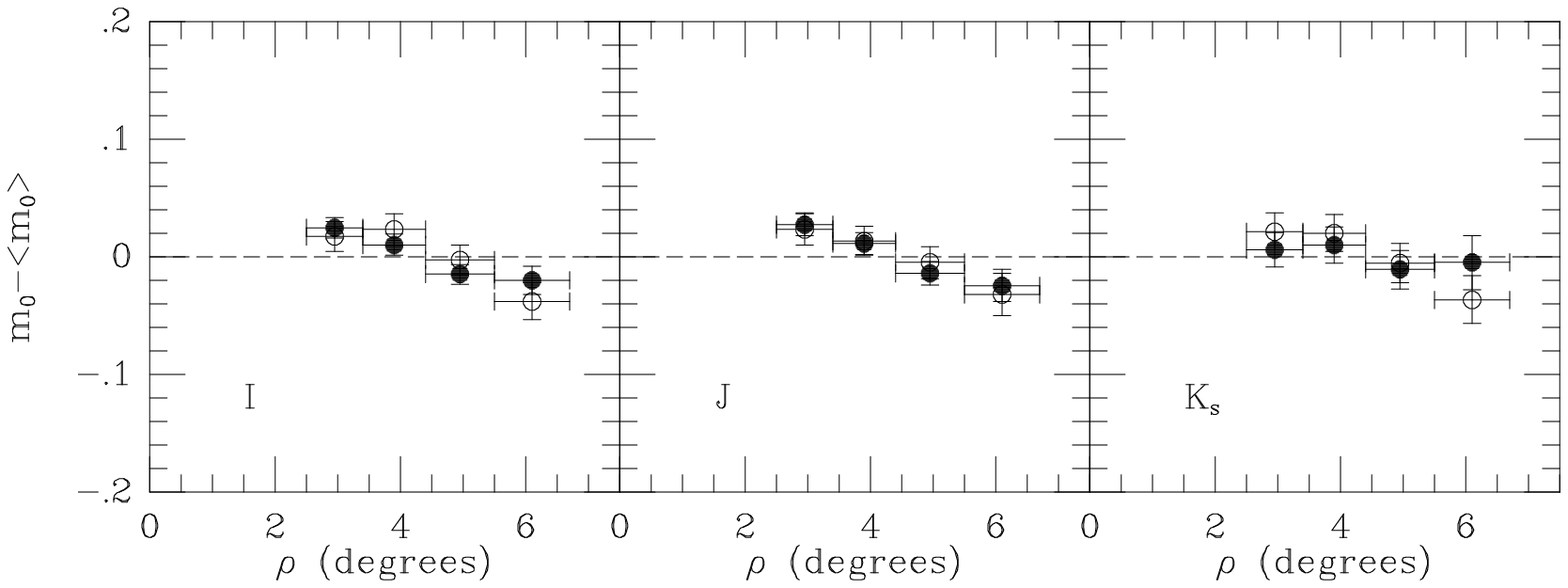}}
\ifsubmode
\vskip3.0truecm
\addtocounter{figure}{1}
\centerline{Figure~\thefigure}
\else\figcaption{\figcapmzero}\fi
\end{figure}


\clearpage
\begin{figure}
\epsfxsize=0.6\hsize
\centerline{\epsfbox{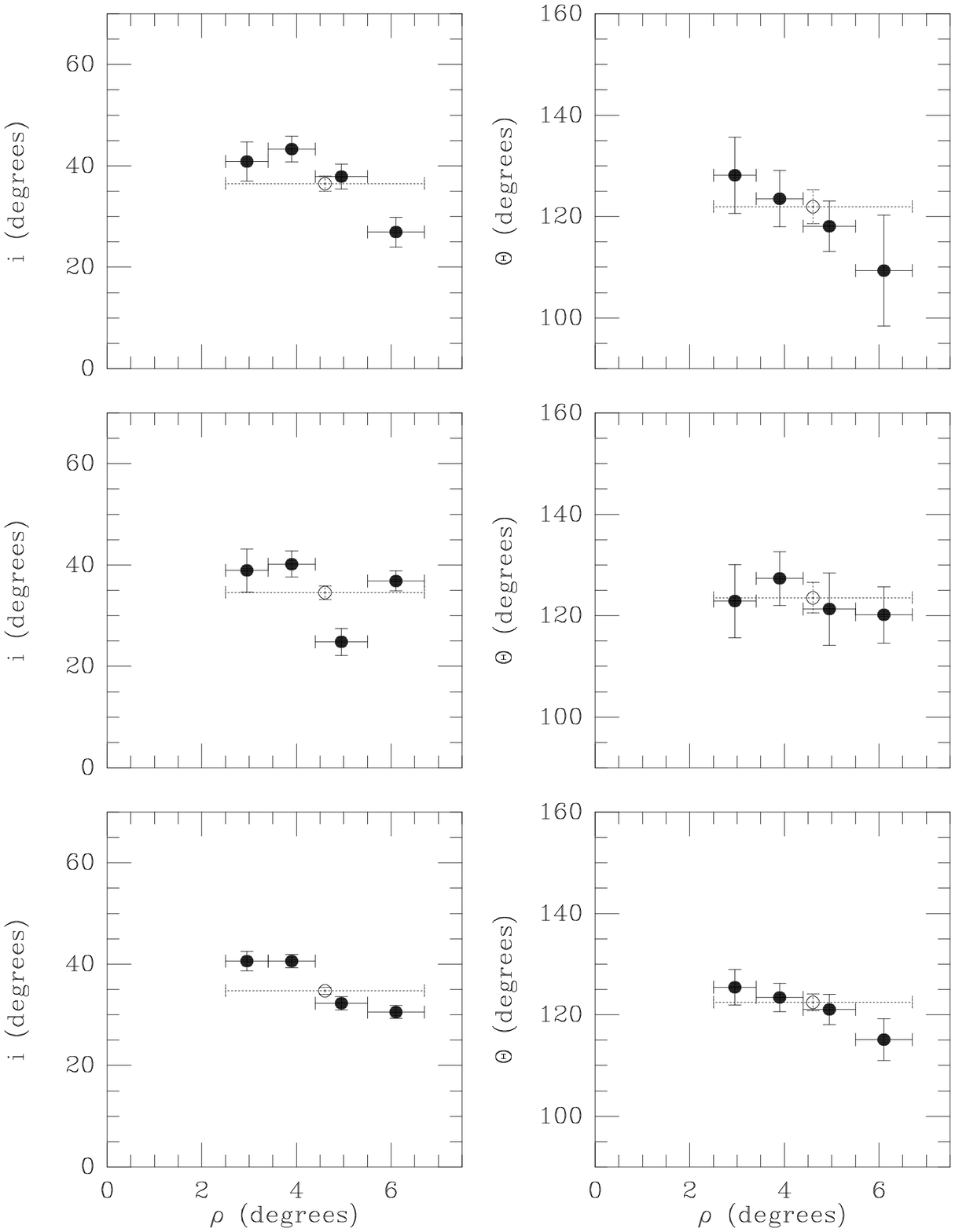}}
\ifsubmode
\vskip3.0truecm
\addtocounter{figure}{1}
\centerline{Figure~\thefigure}
\else\figcaption{\figcapviewang}\fi
\end{figure}


\fi


\clearpage
\ifsubmode\pagestyle{empty}\fi




\end{document}